\documentclass[reprint,aps,prx,superscriptaddress]{revtex4-2}
\usepackage{graphicx}%
\usepackage{amsmath,amssymb,amsfonts}%
\usepackage{amsthm}%
\usepackage{mathrsfs}%
\usepackage{xcolor}%
\usepackage{braket} % braket notation in quantum mechanics
\usepackage{bm} % bold math
\usepackage{tikz} % diagrams in equation
\usepackage[bookmarks=true,colorlinks=true,citecolor=cyan,linkcolor=cyan,filecolor=cyan,urlcolor=cyan]{hyperref} % hyperlink with color
\usepackage{orcidlink}
\usepackage{algorithm}
\usepackage{algpseudocode}
\newcommand{\argmin}{\mathop\mathrm{arg~min}\limits}
\DeclareMathOperator{\Tr}{Tr}

\definecolor{mygreen}{RGB}{0,200,0}

\theoremstyle{plain}
\newtheorem{thm}{Theorem}

\newtheorem{cor}{Corollary}
\newtheorem{lem}{Lemma}

\theoremstyle{definition}
\newtheorem{dfn}{Definition}

\makeatletter
\renewcommand{\ftype@algorithm}{8}
\makeatother

\begin{document}

\title{Markov-Shielded Typicality for Thermal Quantum States}

\author{Atsushi Iwaki\,\orcidlink{0000-0002-5345-8045}}
\email{atsushi-iwaki@phys.s.u-tokyo.ac.jp}
\affiliation{Department of Physics, University of Tokyo, 7-3-1 Hongo, Bunkyo-ku, Tokyo 113-0033, Japan}

\author{Chisa Hotta\,\orcidlink{0000-0001-5898-5963}}
\affiliation{Department of Basic Science, University of Tokyo, 3-8-1 Komaba, Meguro-ku, Tokyo 153-8902, Japan}

\date{\today}

\begin{abstract} 
Pure states of quantum many-body systems at thermal equilibrium are known to be typical, 
in a sense that a randomly chosen pure state from an appropriate energy shell reproduces the Gibbs-state expectation values of 
most observables acting on a small subsystem. 
However, such typical states are intrinsically stochastic, conventionally generated from 
the random high-temperature state, suffer from statistical fluctuations, 
and do not provide a continuous connection to the ground state.
Here, we introduce a deterministic variational framework for constructing a Markov-Shielded Typical (MST) state, 
a unique thermal pure-state representative whose local reduced density matrices closely approximate those of the Gibbs state, 
from the ground state up to infinite-temperature limit. 
Our approach reformulates free-energy minimization by replacing the nonlocal von Neumann entropy with a local, 
computationally tractable Markov entropy, 
where we establish rigorous upper and lower bounds of MST free energy to lie in the very vicinity of 
the thermal free energy. 
This framework extends the concept of typicality to deterministic pure-state representations of thermal equilibrium, 
offering a scalable wave-function approach with potential applications to both classical and quantum computing.
\end{abstract}

\maketitle

%%%%%%%%%%%%%%%%%%%%%%%%%%%%%%%%%%%%%%%%%%%%%%%%%%%%%%%%%%%%%%%%%
\section{Introduction} \label{sec:intro}
The traditional framework of statistical mechanics states that the thermal equilibrium 
is described by an ensemble of mixed states called the Gibbs state. 
Gibbs state is often introduced as a minimizer of the generalized free energy for a Hamiltonian $H$, 
given as a functional of the density operator $\rho$: 
\begin{align}
    \tilde{F}(\rho; T) = \Tr[\rho H] - T ~ S_\mathrm{vN}(\rho), 
    \label{eq:gene_fene}
\end{align}
where $S_\mathrm{vN}(\rho) = - \Tr[\rho\log\rho]$ is the von Neumann entropy. 
At temperature $T = 1/\beta$, the variational principle naturally yields the well-known form of the Gibbs density operator, 
$\rho_\mathrm{G}(\beta) = e^{-\beta H} / Z(\beta)$ with $Z(\beta) = \Tr[e^{-\beta H}]$, 
which gives the thermal entropy $S_\mathrm{th}=S_\mathrm{vN}(\rho_\mathrm{G})$ 
and the thermal free energy $F_\mathrm{th}(T)=\tilde{F}(\rho_\mathrm{G},T)$, accordingly. 

\begin{figure}
\centering
\includegraphics[width=0.9\hsize]{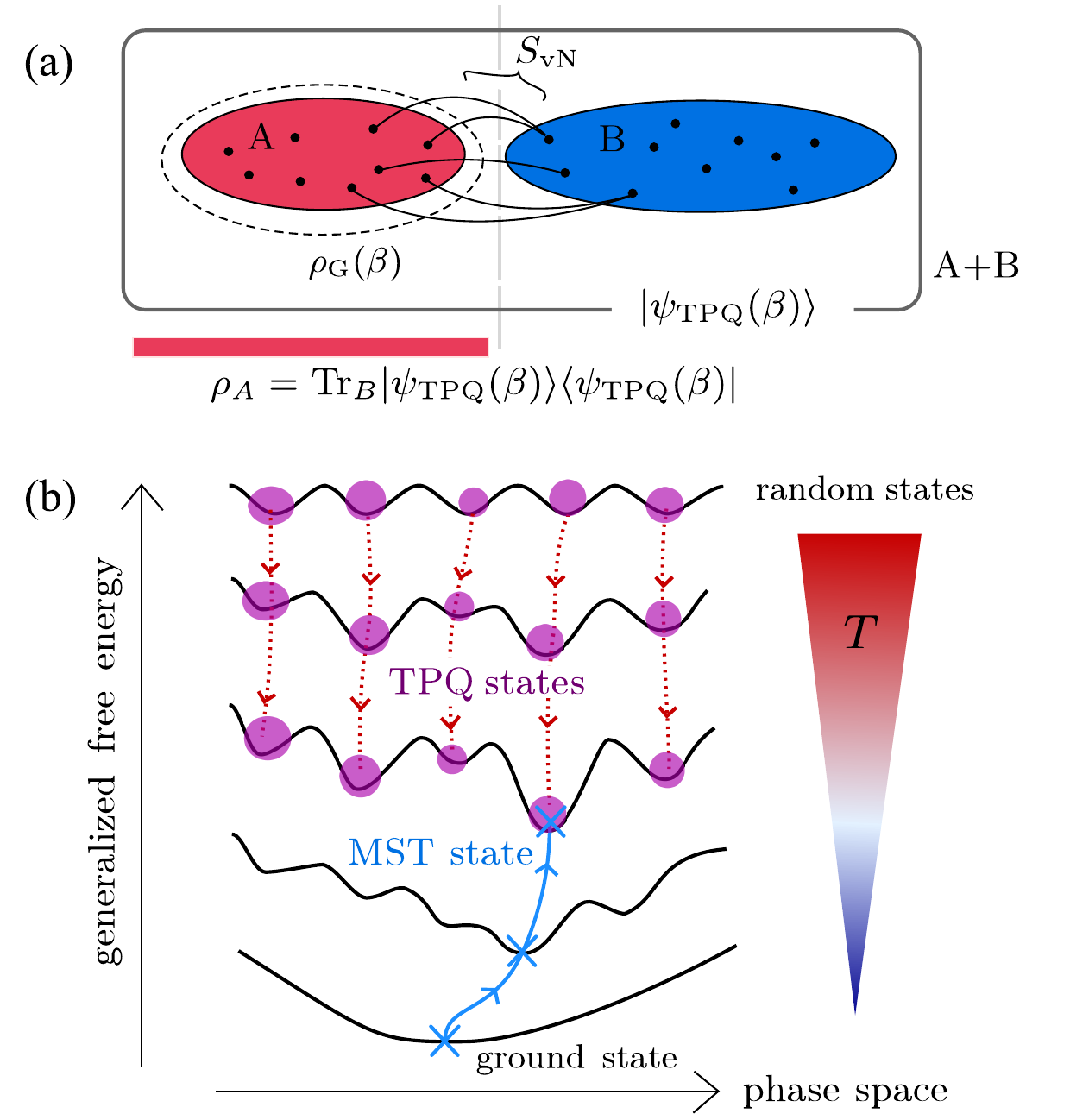}
\caption{
Schematic illustration of (a) TPQ states ($A+B$) and Gibbs state, and (b) the free energy landscape of how the TPQ state evolves from the high temperature limit with imaginary time. Inversely, our MST state starts from $T=0$ and traces the most typical state among the TPQ shells toward the highest temperature. 
}
\label{fig1}
\end{figure}

Modern theories have updated this paradigm to claim that this ensemble operator, 
which costs numbers of averaging that grows exponentially with system size $N$, 
and is often unreachable numerically, is unnecessary, 
and instead one may refer to an individual pure state \cite{popescu2006, goldstein2006}, 
which is later named the thermal pure quantum (TPQ) state
\cite{sugiura2012, sugiura2013, hyuga2014}. 
To clarify their relationships, one needs to compare the open and isolated systems. 
Suppose that the open system $A$ is connected to a bath and equilibrates 
after a long enough timescale to a Gibbs state based on a canonical mixture in connection to a bath. 
On the other hand, if we consider a mixed state in the system $A$ 
and purify it by attaching a sufficient auxiliary state $B$, 
which is proven to be always possible, $A+B$ form an isolated system. 
The eigenstate thermalization hypothesis \cite{deutsch1991, srednicki1994, rigol2008} claims that
the majority of states realized on isolated systems, say $A+B$, in a far-from-equilibrium environment,
can evolve in time to a true equilibrium. 
The philosophy of \textit{canonical typicality}
\cite{popescu2006, goldstein2006, reimann2007, sugita2007} 
asserts that a system $A$, whether embedded in an isolated TPQ state or coupled to a thermal bath, 
exhibits statistically almost indistinguishable reduced properties. 
This equivalence is illustrated schematically in Fig.~\ref{fig1}(a). 
The density operator of $A$ obtained from the pure state of $A+B$ by tracing out the degrees of freedom of $B$, 
$\rho_A = \Tr_B \ket{\psi_\mathrm{TPQ}(\beta)} \bra{\psi_\mathrm{TPQ}(\beta)}$, 
reproduces the mixed-state ensemble with the Boltzmann weight naturally emerging in the process.
An ``overwhelming majority" of TPQ states is thus thermodynamically equivalent to the Gibbs state 
when one considers physical observables within a sufficiently large ``subsystem'' $A$. 

The realization of TPQ state in computers \cite{imada1986, jaklic1994, hams2000}
is demonstrated by utilizing the randomness of unitary matrices, 
which embodies the concept of typical sets. 
Choosing a pure state uniformly at random, $\ket{\psi_0}$, which represents the highest $T$ state, 
the imaginary-time evolution cools the state down to a temperature $\beta^{-1}$ as
$\ket{\psi_\mathrm{TPQ}(\beta)} \propto e^{-\beta H/2} \ket{\psi_0}$. 
Its density matrix, $\rho_\mathrm{TPQ}= {\rm Tr}_B \ket{\psi_\mathrm{TPQ}} \bra{\psi_\mathrm{TPQ}}$, 
agrees with the density matrix of a Gibbs state stochastically at a very high rate. 

However, $\rho_\mathrm{TPQ}$ suffers from two issues by definition. 
First, $\rho_\mathrm{TPQ}$ is a stochastic pure-state representative that has substantial fluctuations 
in finite-size systems that depend on the choice of $\ket{\psi_0}$. 
This fluctuation is proven to be bounded by $e^{- S_\mathrm{th}}$ \cite{sugiura2013},
which increases rapidly at lower temperatures because, intuitively, 
the law of large numbers loses its power, which becomes serious at 
the limited system sizes available for quantum many-body calculations.
This is the implication of the ``overwhelming majority", 
and within that random fluctuation, the states obtained are mostly ``typical.''
Second, it is constructed by cooling 
the highest temperature state $\ket{\psi_0}$ via the imaginary time evolution, which never reaches the ground state. 
Here, accessing low-temperature Gibbs states can be even more challenging, as illustrated by recent studies 
of quasi-local quantum Gibbs samplers \cite{chen2025-2, rouze2026-1, rouze2026-2}. 
These samplers prepare the Gibbs state $\rho_\mathrm{G}$ through Lindblad dissipation, 
and rigorous proofs establish convergence to $\rho_\mathrm{G}$ in polynomial time. 
However, their results are restricted to a high-temperature regime, well above the characteristic energy scale of the model. 
This fact highlights both the importance and the challenge of constructing 
\textit{deterministic pure-state representations of thermal equilibrium over a broad temperature range}.

Then, the question is whether we can prepare an ideal thermal pure state in a computer 
that overcomes the above two issues inherent to TPQ state. 
To qualify how each TPQ state closely resembles the Gibbs state, 
the framework called microscopic thermal equilibrium (MITE) \cite{goldstein2015} was introduced. 
The pure state is said to be MITE if its local density matrices of all subsystems up to a specified size 
called $k$-local (with $k<N/2$) are equal to those of the Gibbs state within an error $\epsilon$. 
In this work, we clarify that a thermal state can be uniquely determined variationally, and mimics MITE. 
To generate such a state, we propose a variational framework  
using a generalized Markov free energy $\tilde{F}_M$ \cite{poulin2011, ferris2013} 
as a \textit{functional of a pure state}, which is 
practically accessible in a quantum many-body calculation for arbitrary temperature. 
Denoting by $\mathcal{S}_{\mathrm{pure}}$ the set of pure-state density matrices, we define
\begin{align}
 \psi_\mathrm{MST}(T) =
 \argmin_{\psi \in \mathcal{S}_{\mathrm{pure}}} \tilde{F}_M(\psi; T).
 \label{eq:mst1}
\end{align}
We call $\psi_\mathrm{MST}(T)=\ket{\psi_\mathrm{MST}(T)}\bra{\psi_\mathrm{MST}(T)}$
a Markov-shielded typical (MST) state. 
Unlike the conventional free energy \eqref{eq:gene_fene} which includes 
the nonlocal von Neumann entropy being strictly zero for the pure state, 
$\tilde{F}_M$ is defined based on local quantities and gives for any subsystem an optimal solution. 
Notably, the ground state, which is unique, continues to Eq.~\eqref{eq:mst1} at finite temperature, 
which makes this MST state the utmost typical and different in philosophy from 
a TPQ state as a stochastic pure-state representative, 
as illustrated in Fig.~\ref{fig1}(b). 

We show analytically that, under physically motivated assumptions,
the Markov free energy of the MST state closely approximates
the thermodynamic free energy, and that approximate MITE states
are nearly optimal candidates for MST optimization. 
Using a tensor network ansatz for a one-dimensional (1D) spin system,
we also demonstrate that the optimized MST states
closely reproduce the local reduced density matrices of the Gibbs state,
providing numerical evidence for the existence of MITE. 

%*%*%*%*%*%*%*%*%*%*%*%*%*%*%*
%*%*%*%*%*%*%*%*%*%*%*%*%*%*%*

%%%%%%%%%%%%%%%%%%%%%%%%%%%%%%%%%%%%%%%%%%%%%%%%%%%%%%%%%%%%%%%%%
\section{Preliminaries}
We consider a quantum many-body system described by a density matrix $\rho$. 
Using its reduced density matrix, 
$\rho^A = \Tr_{A^c}[\rho]$, for a subsystem $A$, 
the entanglement entropy $S(A)_\rho = S_\mathrm{vN}(\rho^A)$ 
is obtained, where we omit the subscript $\rho$ unless otherwise needed. 
If $\rho$ is a thermal equilibrium state, 
then $S(A)_\rho$ is equivalent to the thermal entropy $S_\mathrm{th}$ of $A$,
and we derive a Markov entropy that replaces it. 

\subsection{Quantum Markov property}
\begin{figure}%[tbhp]
\centering
    \includegraphics[width=0.95\hsize]{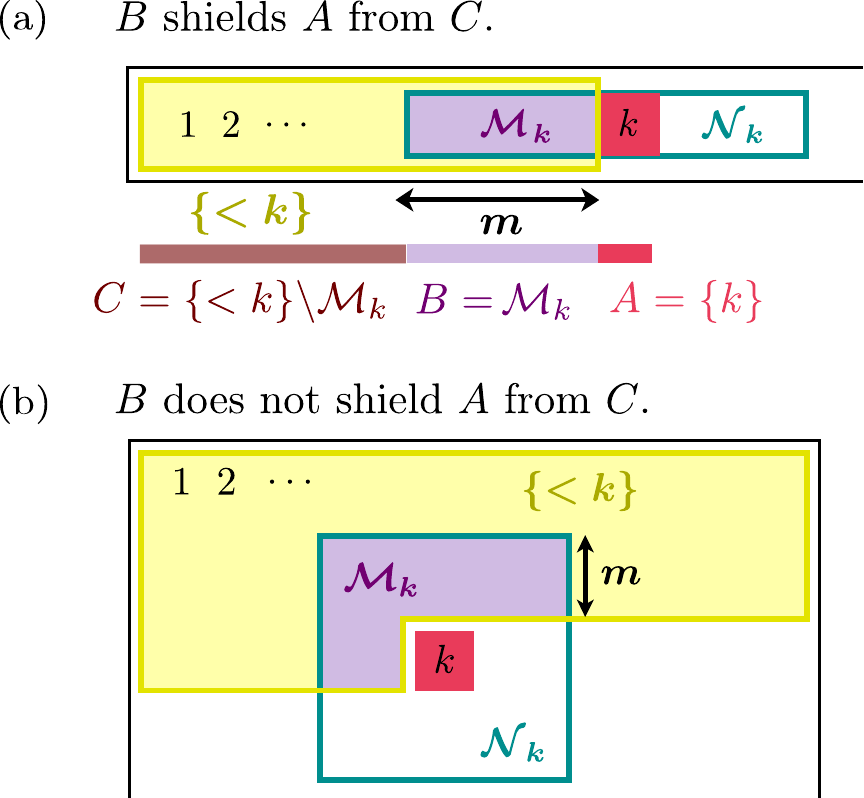}
    \caption{
    In the MED method, lattice sites are labeled according to a specific ordering: 
    (a) in 1D systems, numbering proceeds from left to right, 
    and (b) in 2D systems, from the top left.   
    The Markov shield $\mathcal{M}_k$ (blue) is defined as the intersection between  
    the set of sites labeled with indices smaller than $k$, $\{<k\}$ (yellow),  
    and the neighborhood of site $k$, $\mathcal{N}_k$ (green).  
    The parameter $m$ denotes the diameter of the Markov shield.
    If we take $A = \{k\}$, $B = \mathcal{M}_k$, and $C = \{<k\} \backslash \mathcal{M}_k$,  
    then in case (a), $B$ shields $A$ from $C$, whereas in case (b), it does not.
    }
    \label{fig:markov_shield}
\end{figure}

Let us consider a lattice system and assign a site index by drawing a 1D path. 
We now focus on site $k$, and denote a set of sites labeled by numbers smaller than $k$ 
as $\{ <k \}$, the neighborhood of site $k$ as $\mathcal{N}_k$, 
and $\mathcal{M}_k = \{ <k \} \cap \mathcal{N}_k$ is termed Markov shield. 
We then set the three regions $A, B, C$ as $k$, $\mathcal{M}_k$ and $\{<k\} \setminus \mathcal{M}_k$, respectively, 
to introduce the notion of shielding.
We say that $B$ shields $A$ from $C$ if all paths along the lattice bonds which start in $A$ and end in $C$ go through $B$. 
Figure~\ref{fig:markov_shield} shows two examples that (a) $B$ shields $A$ from $C$, and (b) which does not. 

Remind that in the classical probabilistic theory, 
if a distribution of $X$ is determined solely by $Y$ independent of $Z$, 
it is equivalent to the vanishing of the mutual information 
between $X$ and $Z$ conditioned by $Y$, which is called Markov property, 
as is common in the Monte Carlo simulation. 
The mutual information between $X$ and $Z$ conditioned by $Y$ is defined as 
\begin{align}
    I(X\!:\!Z\!\mid \!Y) = H(XY) + H(YZ) - H(XYZ) - H(Y), 
\end{align}
where $H(X)$ is the Shannon entropy of a random variable $X$. 
Similarly, in quantum systems, by reading off the Shannon entropy as von Neumann entropy, 
the zero of the mutual information for the subsystems $A$, $B$, and $C$, given as 
\begin{equation}
    I(A\!:\!C\!\mid \!B)_\rho = S(AB) + S(BC) - S(ABC) - S(B), 
\end{equation}
implies a quantum Markov property. 
Again in Fig.~\ref{fig:markov_shield}, if the quantum state $\rho$ on a given lattice satisfies 
$I(A\!:\!C\!\mid \!B)_\rho = 0$ for arbitrary choice of $B$ that shields $A$ from $C$, 
it is called a quantum Markov network. 
Indeed, $I(A\!:\!C\!\mid \!B)_\rho=0$ is both a necessary and a sufficient condition to 
recover the information about $A$ using the information solely from $B$, known as the Petz recovery map \cite{petz1986, petz1988}. 
The quantum Markov networks are equivalent to 
the Gibbs states of a commuting Hamiltonian \cite{brown2012}, 
which is the quantum version of the Hammersley-Clifford theorem \cite{hammersley1971}. 
The general Gibbs states have an approximate Markov property
\cite{kato2019, kuwahara2025, kato2025, chen2025, yang2026}. 
In 1D systems, the mutual information is bounded by
\begin{equation}
    I(A\!:\!C\!\mid \!B)_{\rho_\mathrm{G}} \le c_1 e^{-m/\xi}, 
    \label{eq:i_g}
\end{equation}
where $c_1$ and $\xi$ are temperature-dependent constants, 
and $m$ is the size of the Markov shield $B$ \cite{kuwahara2025}.
Here, the polynomial prefactor in the original bound is absorbed into the constants 
by slightly weakening the exponential decay rate.

\subsection{Markov entropy decomposition}
%*%*%*%*%*%*%*%*%*%*%*%*%*%*%*
%*%*%*%*%*%*%*%*%*%*%*%*%*%*%*

We now introduce the Markov entropy decomposition (MED), a method developed and refined to 
provide a lower bound on the thermodynamic free energy \cite{poulin2011, ferris2013}.
Since a pure state has vanishing von Neumann entropy by definition, 
$S_\mathrm{vN}(\ket{\psi}\bra{\psi}) = 0$, 
the direct application of Eq.~\eqref{eq:gene_fene} does not make sense. 
The MED instead allows us to evaluate an upper bound of 
the von Neumann entropy $S_\mathrm{vN}$ using only local quantities. 
We first rewrite it using the conditional entropy $S(A \mid\! B) = S(AB) - S(B)$: 
\begin{align}
    S_\mathrm{vN} (\rho)
    &= S(1) + \sum_{k=2}^N \left[ S(\{\le k\}) - S(\{< k\}) \right] \notag \\
    &= \sum_{k=1}^N S(k \mid \{ <k \}). 
\end{align}
Employing the inequality of strong subadditivity $S(A \mid\! BC) \le S(A \mid\! B)$ \cite{lieb1973}, 
the conditional entropy is bounded by the local quantity 
using a Markov shield: 
\begin{equation}
    S(k \mid \{ <k \}) 
    \le S(k \mid \{ <k \} \cap \mathcal{N}_k) 
    = S(k \mid \mathcal{M}_k). 
\end{equation}
The von Neumann entropy is bounded by their sum as
\begin{equation}
    S_\mathrm{vN}(\rho) 
    = \sum_{k=1}^N S(k \mid \{ <k \}) 
    \le S_M(\rho), 
\end{equation}
where we use the Markov entropy defined as
\begin{align}
    S_M(\rho) = \sum_{k=1}^N S(k \mid \mathcal{M}_k). 
\label{eq:sm}
\end{align}
The generalized free energy Eq.~\eqref{eq:gene_fene} is then bounded 
by the generalized Markov free energy $\tilde{F}_M$ as 
\begin{equation}
    \tilde{F}_M(\rho; T) = \Tr[\rho H] - T ~ S_M(\rho) \le \tilde{F}(\rho; T). 
    \label{eq:F_M}
\end{equation}
Since Eq.~\eqref{eq:F_M} holds for any given $\rho$, 
the inequality is preserved upon minimizing both sides with respect to $\rho$. 
The minimization of the generalized Markov free energy $\tilde{F}_M (\rho;T)$ 
gives the Markov free energy $F_M(T)$, and we find, 
\begin{equation}
F_M(T) = \min_\rho \tilde{F}_M(\rho; T)
\le \min_\rho \tilde{F}(\rho; T) = F_\mathrm{th}(T). 
\label{eq}
\end{equation}
This means that $F_M(T)$ gives a lower bound for the thermodynamic free energy $F_\mathrm{th}(T)$. 
\par
In numerical calculations, the MED algorithm optimizes a set of local density matrices subject 
to consistency constraints on their overlapping regions \cite{poulin2011, ferris2013}. 
Although these constraints ensure consistency among overlapping marginals, 
they do not necessarily guarantee the existence of a global \textit{physical} state 
whose reduced density matrices reproduce all the optimized marginals. 
This reflects the intrinsic difficulty of the quantum marginal problem \cite{liu2006, schilling2014}. 
Nevertheless, the MED algorithm is rigorously guaranteed to provide a lower bound on the thermal free energy. 

%%%%%%%%%%%%%%%%%%%%%%%%%%%%%%%%%%%%%%%%%%%%%%%%%%%%%%%%
%%%%%%%%%%%%% Markov-shielded typicality %%%%%%%%%%%%%%%
%%%%%%%%%%%%%%%%%%%%%%%%%%%%%%%%%%%%%%%%%%%%%%%%%%%%%%%%
\section{Theoretical framework} \label{sec:theory}

We consider a 1D quantum lattice system consisting of $N$ sites 
with open boundary conditions (OBC) 
and a local degrees of freedom $d$ at each site, 
described by a finite-range Hamiltonian 
\begin{equation}
    H = \sum_{i} h_i,
\end{equation}
where $h_i$ denotes a local interaction term belonging to site $i$.
We assume that the number of $h_i$ is formally smaller than or equal to $N$ 
and that their operator norms are bounded by 
\begin{equation}
    J = \max_i \|h_i\|_\mathrm{op},
\end{equation}
for some finite $J$ 
where $\|\cdot\|_\mathrm{op}$ denotes the operator norm. 
%
%*%*%*%*%*%*%*%*%*%*%*%*%*%*%*
\begin{figure}%[tbhp]
\centering
    \includegraphics[width=0.95\hsize]{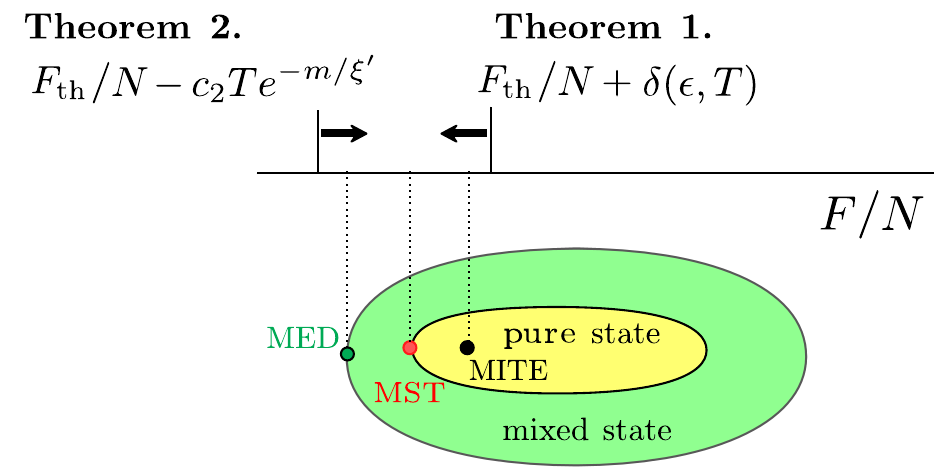}
    \caption{Schematic illustration of our proofs in Sec.\ref{sec:mstproof}. 
    From Theorem~\ref{thm:mite}, we find 
    $\tilde F_M(\psi_\mathrm{MST})/N\le \tilde F_M(\psi_\mathrm{MITE})/N 
    \le \tilde F_M(\rho_\mathrm{G})/N +\delta(\epsilon, T) \le F_\mathrm{th}/N + \delta(\epsilon, T)$. 
    Similarly, from Theorem~\ref{thm:med}, 
    $F_\mathrm{th}/N -c_2T e^{-m/\xi'} \le  \tilde F_M(\rho_\mathrm{MED})/N \le  \tilde F_M(\psi_\mathrm{MST})/N$. 
    These inequalities guarantee that the MST free energy is bounded from both directions by $F_\mathrm{th}$. 
    }
    \label{fig:freeenergies}
\end{figure}
%*%*%*%*%*%*%*%*%*%*%*%*%*%*%*
%
\subsection{Markov-shielded typical state}
\label{sec:mstproof}
We now pin down the relationships between the thermal free energy $F_\mathrm{th}$ minimized via Eq.~\eqref{eq:mst1} 
and the other free energies about the mixed and pure states, 
which are schematically illustrated in Fig. \ref{fig:freeenergies}. 
The Markov free energy of the MST state lies in between the lower and upper bounds 
of $F_\mathrm{th}$, mediated by the inequalities between the free energies of the MED and MITE, 
which we formulate in the following. 
\par
We consider the setup illustrated in Fig.~\ref{fig:markov_shield}(a), where $B$ shields $A$ from $C$. 
Based on Eq.~\eqref{eq:F_M}, the difference between the two free energies is given by the conditional mutual information as
\begin{align}
    \tilde{F}(\rho; T) - \tilde{F}_M(\rho; T)
    &= T \sum_{k=1}^N [ S(k \mid \mathcal{M}_k)-S(k \mid \{ <k \})] \notag \\
    &= T \sum_{k=1}^N I(k:\{ <k \} \backslash \mathcal{M}_k \mid \mathcal{M}_k). 
    \label{eq:fdiff}
\end{align}
Plugging $\rho_\mathrm{G}$ into Eq.~\eqref{eq:fdiff} 
and combining it with Eq.~\eqref{eq:i_g}, we obtain 
\begin{equation}
    0 \le \frac{F_\mathrm{th}(T) - \tilde F_M(\rho_\mathrm{G};T)}{N} \le c_1 T e^{- m / \xi}, 
    \label{eq:fdiff_g}
\end{equation}
where we used $\tilde{F}(\rho_\mathrm{G}; T)=F_\mathrm{th}(T)$. 
This shows that the Markov free energy density of the Gibbs state 
approaches the thermal free energy density exponentially 
with increasing Markov shield size $m$.
\par
Next, we introduce the pure state that approximates MITE \cite{goldstein2015}. 
\begin{dfn}
A pure state $\psi_\mathrm{MITE} \in \mathcal{S}_{\mathrm{pure}}$
is said to be an $\epsilon$-approximate $k$-local MITE state if 
\begin{equation}
    D(\psi_\mathrm{MITE}^A,\rho_{\mathrm{G}}^A) \le \epsilon
\end{equation}
for every contiguous subsystem $A$ with $|A|\le k$,
where $D(\rho,\sigma)=\Tr[|\rho-\sigma|]/2$ denotes the trace distance.
\end{dfn}
\noindent
Hereafter, we refer to such a state as an $(\epsilon,k)$-MITE state for brevity.
The following lemma bounds the difference between the Markov free energies
of an $(\epsilon,m+1)$-MITE state and the Gibbs state.
In what follows, we assume that each local Hamiltonian term
is supported within at most $m+1$ consecutive sites.
\begin{lem} \label{lem:mite_fene}
For any $(\epsilon,m+1)$-MITE state $\psi_\mathrm{MITE}$, 
\begin{equation}
\frac{|\tilde{F}_M(\psi_\mathrm{MITE};T)-\tilde{F}_M(\rho_\mathrm{G};T)|}{N} \le \delta(\epsilon, T), 
\end{equation}
where
\begin{align}
    \delta(\epsilon, T)
    &=2J\epsilon+
    T\left[2\epsilon\log d+(1+\epsilon)h_2\left(\frac{\epsilon}{1+\epsilon}\right)\right] \notag \\
    &= \mathcal{O}(\epsilon T \log (1/\epsilon)) \quad 
    (T>0, ~ \epsilon \to 0),
\end{align}
and $h_2(x)=-x\log x-(1-x)\log(1-x)$ is the binary entropy.
\end{lem}
\begin{proof}
By the definition of $(\epsilon,m+1)$-MITE, the energy difference satisfies
\begin{align}
    \left|\Tr[(\psi_\mathrm{MITE}-\rho_\mathrm{G})H]\right|
    &\le 2\sum_i \|h_i\|_\mathrm{op}
    D(\psi_\mathrm{MITE}^{A_i},\rho_\mathrm{G}^{A_i}) \notag\\
    &\le 2NJ\epsilon.
\end{align}
For the entropy term, 
the continuity bound for quantum conditional entropy
\cite{winter2016} gives
\begin{align}
    &|S(k \mid \mathcal{M}_k)_{\psi_\mathrm{MITE}}
    -S(k \mid \mathcal{M}_k)_{\rho_\mathrm{G}}| \notag\\
    &\le 2\epsilon\log d
    +(1+\epsilon)h_2\left(\frac{\epsilon}{1+\epsilon}\right).
\end{align}
Summing over all sites and combining this with the energy bound
yields the stated result.
\end{proof}

If an $(\epsilon,m+1)$-MITE state exists and $\delta(\epsilon,T)$ is small,
its Markov free energy and that of the MST state
provide approximate lower bounds on the thermodynamic free energy.
Combining the previous lemma with Eq.~\eqref{eq:fdiff_g}
and the MST variational principle, we obtain the following theorem.
\begin{thm} \label{thm:mite}
Assuming the existence of an $(\epsilon,m+1)$-MITE state $\psi_\mathrm{MITE}$, 
we have
\begin{equation}
    \frac{\tilde F_M(\psi_\mathrm{MITE};T)-F_\mathrm{th}(T)}{N}
    \le \delta(\epsilon,T).
\end{equation}
By the variational principle in Eq.~\eqref{eq:mst1},
the same bound holds for $\psi_\mathrm{MST}$.
\end{thm}
\begin{proof}
From Eq.~\eqref{eq:fdiff_g}, we have
\begin{equation}
    \tilde F_M(\rho_\mathrm{G};T) \le F_\mathrm{th}(T).
\end{equation}
Combining this with Lemma~\ref{lem:mite_fene}, we obtain
\begin{align}
    \tilde F_M(\psi_\mathrm{MITE};T)
    &\le \tilde F_M(\rho_\mathrm{G};T)
    +N\delta(\epsilon,T) \notag\\
    &\le F_\mathrm{th}(T)+N\delta(\epsilon,T),
\end{align}
which yields the stated bound.
By the variational definition of the MST state in Eq.~\eqref{eq:mst1},
\begin{equation}
    \tilde F_M(\psi_\mathrm{MST};T)
    \le \tilde F_M(\psi_\mathrm{MITE};T),
\end{equation}
so the same bound holds for $\psi_\mathrm{MST}$.
\end{proof}
\noindent
This theorem now proves that $F_\mathrm{th}$ gives the upper bound of $\tilde F_M(\psi_\mathrm{MITE})$, 
which is the free energy of the pure state mimicking the Gibbs state 
and satisfying at most the canonical typicality condition. 
\par
The next step is to show that $F_\mathrm{th}$ bounds the pure state free energy from the opposite side. 
For that purpose, we introduce the MED state which is defined as a minimizer of the Markov free energy 
over all physical density matrices: 
\begin{equation}
    \rho_\mathrm{MED}
    = \argmin_{\rho} \tilde{F}_M(\rho;T).
\end{equation}
We assume that $\rho_\mathrm{MED}$ satisfies
the approximate Markov property
\begin{equation}
    I(A\!:\!C\!\mid\!B)_{\rho_\mathrm{MED}}
    \le c_2 e^{-m/\xi'},
    \label{eq:i_med}
\end{equation}
for every tripartition associated with a Markov shield 
as shown in Fig.~\ref{fig:markov_shield}(a), 
where $c_2$ and $\xi'$ is temperature-dependent constants. 
Under this assumption, the Markov free energy of the MST state
provides an approximate upper bound on the thermodynamic free energy
when $m$ is sufficiently large. 
\begin{thm} \label{thm:med}
Assuming the approximate Markov property of the MED state in Eq.~\eqref{eq:i_med}, 
the MST state satisfies
\begin{equation}
    \frac{F_\mathrm{th}(T) - \tilde{F}_M(\psi_\mathrm{MST};T)}{N}
    \le c_2 T e^{-m/\xi'}. 
\end{equation}
\end{thm}
\begin{proof}
By the approximate Markov property of $\rho_\mathrm{MED}$
and Eq.~\eqref{eq:fdiff}, we have
\begin{align}
    0 &\le \tilde{F}(\rho_\mathrm{MED};T)
    -\tilde{F}_M(\rho_\mathrm{MED};T) \notag\\
    &\le c_2 NT e^{-m/\xi'}.
\end{align}
Since $F_\mathrm{th}(T)\le\tilde{F}(\rho_\mathrm{MED};T)$,
it follows that
\begin{equation}
    \tilde{F}_M(\rho_\mathrm{MED};T)
    \ge F_\mathrm{th}(T)-c_2 NT e^{-m/\xi'}.
\end{equation}
By the variational definitions of the MED and MST states,
\begin{equation}
    \tilde{F}_M(\rho_\mathrm{MED};T)
    \le \tilde{F}_M(\psi_\mathrm{MST};T).
\end{equation}
Combining these inequalities yields the bound stated.
\end{proof}
\par
The above two theorems give the following corollary: 
\begin{cor} \label{cor:mst_bounds}
Assuming the existence of an $(\epsilon,m+1)$-MITE state $\psi_\mathrm{MITE}$
and the approximate Markov property of the MED state in Eq.~\eqref{eq:i_med}, 
the MST state satisfies
\begin{equation}
    -c_2 T e^{-m/\xi'}
    \le
    \frac{\tilde{F}_M(\psi_\mathrm{MST};T)-F_\mathrm{th}(T)}{N}
    \le \delta(\epsilon,T).
\end{equation}
For any such MITE state,
\begin{align}
    0 &\le
    \frac{\tilde{F}_M(\psi_\mathrm{MITE};T)
    -\tilde{F}_M(\psi_\mathrm{MST};T)}{N} \notag\\
    &\le \delta(\epsilon,T)+c_2 T e^{-m/\xi'}.
\end{align}
\end{cor}
\noindent
When both $\delta(\epsilon,T)$ and $c_2 T e^{-m/\xi'}$ are small or approaches zero,
the MST Markov free energy approximates the thermodynamic free energy,
and approximate MITE state becomes the optimal candidate
for MST optimization. 
These results provide a theoretical basis for the MST variational principle in Eq.~\eqref{eq:mst1}, 
where the MST serves as the most typical thermal equilibrium state, 
as far as the minimum $\tilde F_M$ is safely identified. 
We notice that the exact MITE states with $\epsilon=0$ have been explicitly constructed 
at infinite temperature \cite{doroudiani2020, wang2021, chiba2024, mohapatra2025, hokkyo2026}, 
but not at $T<\infty$, meaning that our construction and numerical demonstration 
together provide evidence for their existence.

%%%%%%%%%%%%%%%%%%%%%%%%%%%%%%%%%%%%%%%%%%%%%%%%%%%%%%%%%%%%%%%%%
\subsection{MST algorithm} \label{sec:method}
Supported by the proof in the previous subsection, 
we propose a numerical approach to obtain the MST state variationally. 
Here, any types of description of the wave function $\ket{\psi(\bm{\theta})}$ using a finite 
number of variational parameters $\bm{\theta}$ can be used, 
such as variational Monte Carlo methods \cite{mcmillan1965}, 
Boltzmann machine \cite{carleo2017}, 
and density matrix renormalization group \cite{white1992, white1993}. 
\par
The optimization process involves minimizing the Markov free energy with respect to the variational parameters $\bm{\theta}$.
One needs to straightforwardly optimize the wave function by using 
the Markov free energy $\tilde{F}_M(\ket{\psi} \bra{\psi}, T)$ as a complex cost function 
for each given $T$. 
However, because $\tilde{F}_M$ is a functional of $\bm{\theta}$ with many local optima, 
we start from the ground state that minimizes the Hamiltonian, 
and then using the solution $\ket{\psi(\bm{\theta}_T)}$ at $T$ as an initial state, 
obtain $\ket{\psi(\bm{\theta}_{T+\Delta T})}$, by gradually raising the temperature up to $T_\mathrm{max}$. 
The temperature-continuation procedure is summarized in Algorithm~\ref{alg:mst}. 

\begin{algorithm}[t]
\caption{Variational optimization of MST states}
\label{alg:mst}
\begin{algorithmic}[1]
\Require Hamiltonian $H$,
variational wave function $\ket{\psi(\bm{\theta})}$,
Markov free energy $\tilde{F}_M$,
temperature grid $0=T_0<T_1<\cdots<T_L=T_{\max}$
\Ensure Optimized variational states
$\{\ket{\psi(\bm{\theta}_\ell)}\}_{\ell=0}^{L}$
\State Initialize $\bm{\theta}_0$ by approximately
minimizing the ground-state energy
\For{$\ell=1,\ldots,L$}
    \State Set
    $\bm{\theta}\gets\bm{\theta}_{\ell-1}$
    \State Minimize
    $\tilde{F}_M(
    \ket{\psi(\bm{\theta})}
    \bra{\psi(\bm{\theta})};T_\ell)$
    with respect to $\bm{\theta}$
    \State Store the optimized parameters as
    $\bm{\theta}_\ell$
\EndFor
\State \Return
$\{\ket{\psi(\bm{\theta}_\ell)}\}_{\ell=0}^{L}$
\end{algorithmic}
\end{algorithm}

Although computing the ground state is classified as QMA-complete \cite{kitaev2002, aharonov2009, hallgren2013}, 
implying that it is generally hard, most of the physical models with short-range interactions, 
do not apply to these worst-case scenarios. For instance, by focusing on frustration-free systems, where the ground state can be obtained easily, 
we can circumvent this problem. 

We also notice the crucial difference between our MST and MED algorithm. 
The MED algorithm optimizes the locally consistent density matrices, 
whereas our approach considers the Markov free energy $\tilde F_M$ as a functional of a global \textit{pure} state $\psi$. 
The phase space to be optimized also differ; 
the MED optimization problem is always convex 
because it is formulated in terms of local density operators. 
The MST state is defined over the pure state manifold, $\psi \in \mathcal{S}_\mathrm{pure}$ in Eq.~\eqref{eq:mst1}, 
which makes the optimization problem nonconvex. 
We demonstrate in our simulation that we are indeed able to reach the minimum 
by continuously tracking the variational solution from the ground state as the temperature increases. 
%%%%%%%%%%%%%%%%%%%%%%%%%%%%%%%%%%%%%%%%%%%%%%%%%%%%%%
%%%%%%%%%%%%% numerical demonstrations %%%%%%%%%%%%%%%
%%%%%%%%%%%%%%%%%%%%%%%%%%%%%%%%%%%%%%%%%%%%%%%%%%%%%%

\section{Numerical demonstrations} \label{sec:numerical}

\subsection{MST-MPS implementation}

For the numerical demonstration, we employ a matrix product state (MPS) ansatz \cite{fannes1992} 
to optimize the Markov free energy in $\mathcal{S}_\mathrm{pure}$. 
Since the standard MPS-form suffers a severe area-law bound of the entanglement entropy in practice, 
we adopt the representation called TPQ-MPS with auxiliary degrees of freedom attached to both ends of the chain, 
which we have previously succeeded in recovering the volume-law entanglement entropy 
for the description of the thermal state \cite{iwaki2021}:
\begin{align}
    \ket{\psi(\bm{\theta})}
    &= \sum_{\{\alpha\}}^\chi \sum_{\{i\}}^d
    A_{\alpha_0 \alpha_1}^{[1]i_1} A_{\alpha_1 \alpha_2}^{[2]i_2} \cdots
    A_{\alpha_{N-1} \alpha_N}^{[N]i_N} \notag \\
    &\quad\quad\quad\quad\quad \times 
    \ket{\alpha_0, i_1, i_2, \dots , i_N, \alpha_N} \notag \\ 
    &= 
    \raisebox{-0.5\height}{
      \begin{tikzpicture}[x=3mm,y=3mm]
         \draw[line width=1.5pt,rounded corners=7pt]
            (0,-2)--(0,0)--(7,0);
         \draw[line width=1.5pt,rounded corners=7pt]
            (9,0)--(13,0)--(13,-2);
         \foreach \x in {7.5, 8, 8.5}{
            \draw[fill=black] (\x, 0) circle [radius=0.05];
         }
         \draw (2,0)--(2,-2);
         \draw (5,0)--(5,-2);
         \draw (11,0)--(11,-2);
         \draw[fill=red!30] (2,0) circle (1);
         \draw[fill=red!30] (5,0) circle (1);
         \draw[fill=red!30] (11,0) circle (1);
      \end{tikzpicture} 
   } 
\end{align}
which we have previously developed for calculating the so-called TPQ-MPS \cite{iwaki2021}. 
The advantage of this form is that it could maximize the amount of entanglement 
stored in the pure state, allowing us to have a volume-law entanglement entropy 
when measured from the center sites. 
This time, we use it as a variational ansatz and call it MST-MPS. 
The bond dimension $\alpha_0$ and $\alpha_N$ are set to 1 for the ground state, while it increases 
naturally when we increase $T$, due to the increasing entanglement. When starting at finite $T$ one needs to set 
$\alpha_0=\alpha_N=\chi$ the same as other bond dimensions 
(see Supplementary information \cite{supple}). 

The variational process consists of updating the $n$-th tensor given below 
by setting the orthogonal center to that tensor to keep its canonical form: 
\begin{align}
    \raisebox{-0.5\height}{
      \begin{tikzpicture}[x=3mm,y=3mm]
         \draw[line width=1.5pt]
            (0,0)--(4,0);
         \draw (2,0)--(2,-2);
         \draw[fill=red!30] (2,0) circle (1);
      \end{tikzpicture} 
   }
   = \sum_{\alpha\beta}^\chi \sum_{i_n}^d A_{\alpha\beta}^{[n]i_n} \ket{\alpha, {i_n}, \beta}, 
\end{align}
and a set of processes of shifting $n$ from the left edge to the right edge and then back to the left, 
constitutes a single sweep of the update. 
At each step, the $n$-th tensor is optimized to maximize the Markov entropy, 
where its contribution is mediated through the reduced density matrices entering 
the $k=1:N$-th $(m+1)$-local region of $S(k|{\cal M}_k)$ in Eq.~\eqref{eq:sm}, 
which is connected to the $n$-th tensor via the virtual bonds 
$\alpha$ or $\beta$. 
Although the variational parameters are the local $n$-th tensor,
and this update does not change the remaining $N-1$ tensors themselves, 
all the other marginals (local density matrices) change simultaneously, 
making this process nonlocal. 
At the same time, the whole set of MPS retains its pure state property by construction, 
namely all marginals automatically preserve the consistency condition 
as part of a single pure state. 

This is in sharp contrast to the numerical implementation of the MED; 
the standard formulation is such that a single local density operator (marginal) 
is updated in a way to satisfy consistency constraints 
with its neighboring marginals that have overlaps. 
However, this does not imply that all these marginals are part of the same 
global quantum pure state in a compatible manner. 
Such global compatibility is computationally too demanding. 

We use \texttt{ITensors.jl} for the MPS calculations \cite{itensor1, itensor2} 
and \texttt{Optim.jl} for the BFGS optimization of each tensor \cite{mogensen2018}. 
The system size and bond dimension were fixed at $N = 48$ and $\chi = 8$, respectively.  
We set $\Delta T = 0.1$ and $T_\mathrm{max} = 2.0$, 
and at each temperature, we perform up to maximally 10 sweeps, 
while setting the convergence of $F_M / N$ between sweeps to be less than $10^{-4}$. 
The computational bottleneck of this procedure 
is the diagonalization of a series of $(m+1)$-site reduced density operators 
to evaluate the Markov entropy.  
The cost of optimizing a single matrix scales as $\mathcal{O}(N d^{3m})$, 
and a full sweep through the chain costs $\mathcal{O}(N^2 d^{3m})$, 
where $d$ is the local degree of freedom. 
For $m = 4$, the computation time per sweep took about 3 hours for a standard workstation.
\par
In the following demonstrations, 
we use the 1D transverse-field Ising model with OBC, 
\begin{equation}
    H = - \sum_{i=1}^{N-1} \sigma_i^z \sigma_{i+1}^z
        -g \sum_{i=1}^N \sigma_i^x,
\end{equation} 
where $\sigma_i^{\alpha}$ ($\alpha = x, y, z$) denotes the Pauli operator at site $i$.

\subsection{Escape from random fluctuations}

%*%*%*%*%*%*%*%*%*%*%*%*%*%*%*
\begin{figure}%[tbhp]
\centering
    \includegraphics[width=1.0\hsize]{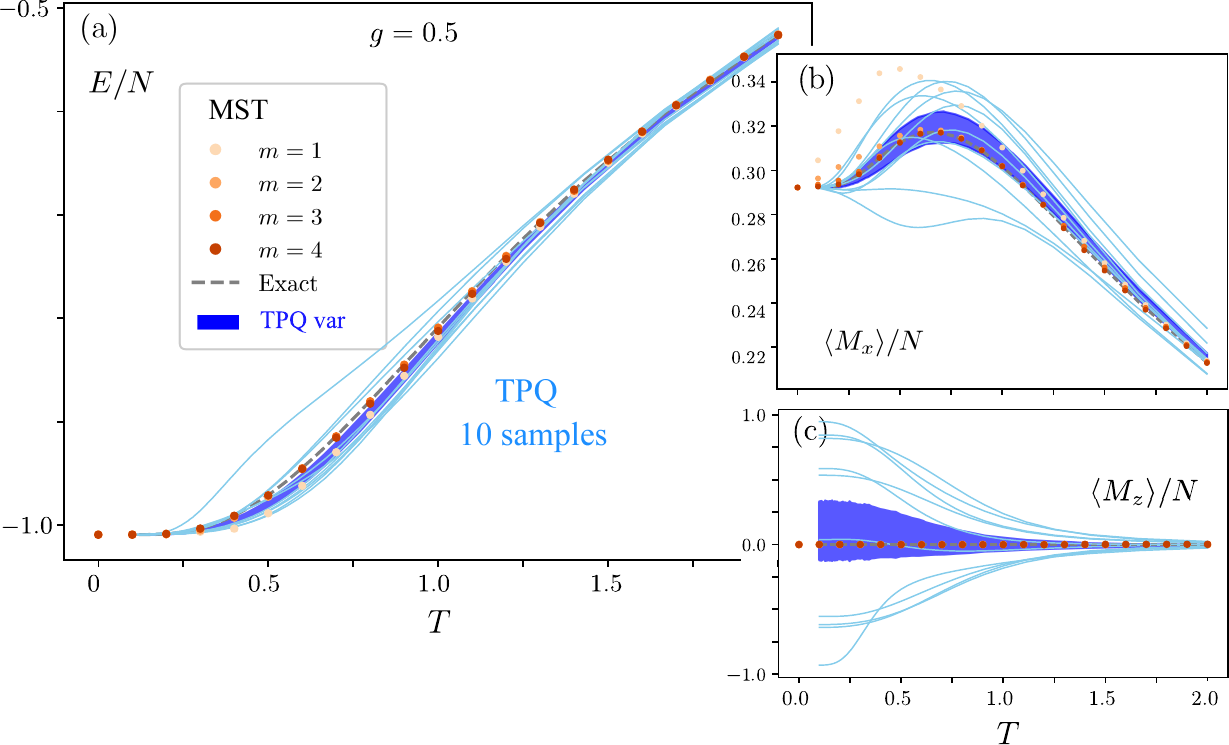}
    \caption{
    Comparison of the one-shot calculations of the TPQ state obtained 
    by the imaginary time evolution using the full Hilbert space (blue lines), 
    and the MST states with $m=1,2,3,4$ using the MPS ansatz. 
    (a) Energy, (b) $x$ and (c) $z$ magnetization of the transverse-field Ising chain with $g=0.5$ for $N=16$. 
    Solid blue lines are the 10 different samples of the TPQ state, blue region is their variance, 
    and the broken line is the exact result. 
    }
    \label{fig:randomfluc}
\end{figure}
%*%*%*%*%*%*%*%*%*%*%*%*%*%*%*

We first compare the MST state obtained using the MST-MPS ansatz 
with TPQ states generated in the full Hilbert space, 
focusing on a small system of size $N=16$. 
We plot in Fig.~\ref{fig:randomfluc}(a-c) the energy $E = \langle H \rangle$, and magnetization 
$M_\alpha = \sum_{i=1}^N \sigma_i^\alpha$ in $\alpha=x$ and $z$ directions. 
For the TPQ state, we plot 10 different samples for the initial choice of random state, $\ket{\psi_0}$, 
and for the MST, only one sample as data points, which we tested for the shield of $m=1,2,3,4$. 
One finds that at $m=3$, the curve almost falls onto the exact diagonalization curve. 
Whereas, the standard TPQ state cannot get rid of the large random fluctuation. 
Notice that this random fluctuation is not the numerical error but the outcome of 
typicality, and is reduced when the system size becomes larger due to the entropic effect, 
but even at relatively large $N$, it largely affects the data at low temperature. 
Picking up a single TPQ state thus provides an increasingly inaccurate result 
at low temperatures, where statistical fluctuations are more pronounced.
This proves the physical picture shown in Fig.~\ref{fig1} that 
the MST can track the most desirable curve \textit{by a one-shot calculation}, 
which, for the TPQ state, is the curve obtained after taking a number of random sampling averages. 

\subsection{Numerical accuracy}

%*%*%*%*%*%*%*%*%*%*%*%*%*%*%*
\begin{figure*}
    \centering
    \includegraphics[width=0.95\hsize]{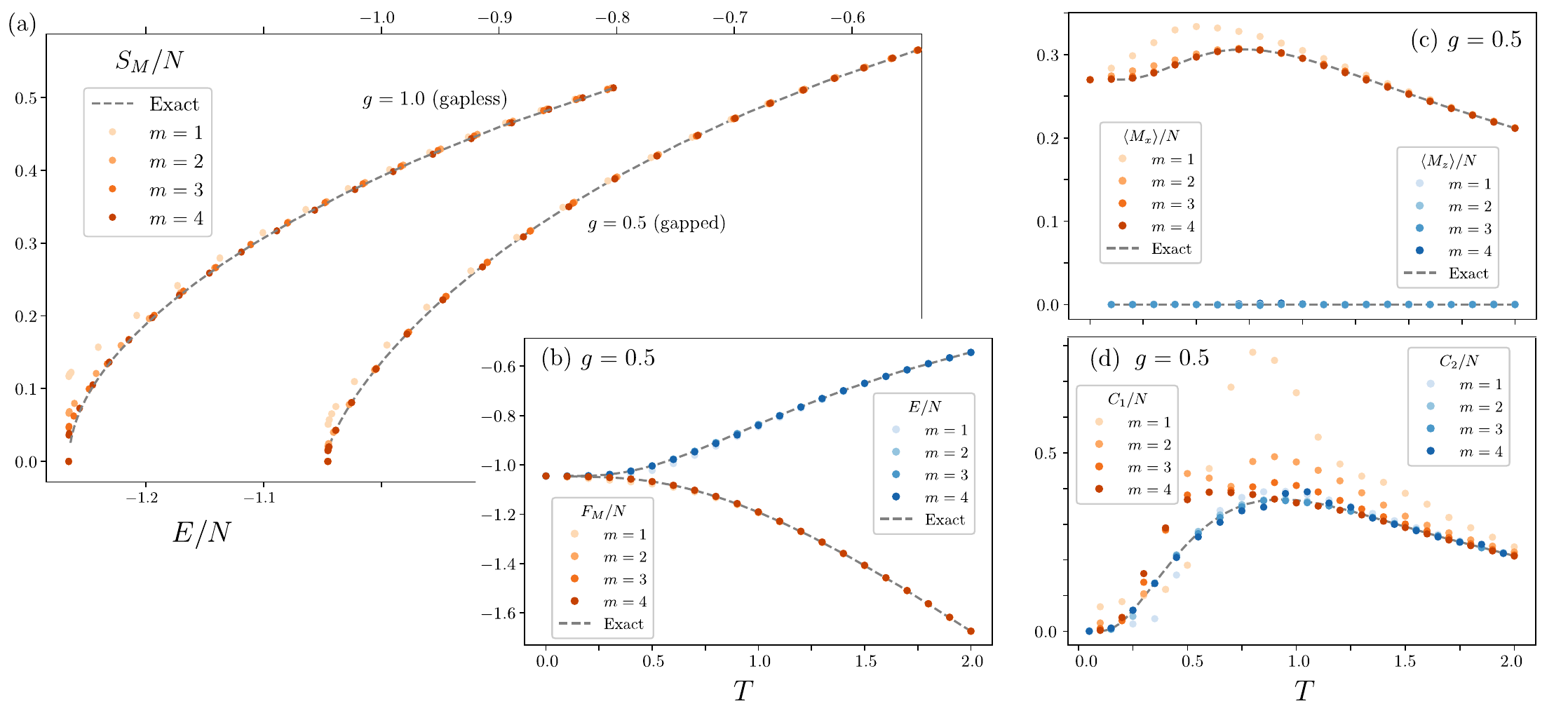}
    \caption{
    Physical quantities of the transverse-field Ising chain for $N=48$. 
    We use the MST-MPS method 
    for various sizes of Markov shields $m = 1, 2, 3, 4$ 
    and the purification method \cite{feiguin2005} 
    as the exact results (dashed lines). 
    (a) $S_\mathrm{M}/N$ as a function of $E/N$ for $g=0.5$ and $1$.
    Temperature dependence of 
    (b) Markov free energy $F_M/N$ and energy $E/N$, 
    (c) transverse magnetization $\langle M_x \rangle / N$, 
    and longitudinal magnetization $\langle M_z \rangle / N$, 
    (d) specific heat per site calculated in two different ways: 
    $C_1$ from Hamiltonian fluctuations, and $C_2$ from temperature derivatives of energy, 
    calculated for $g=0.5$. 
    For the critical case at $g=1.0$, see Supplementary Information \cite{supple}.
    }
    \label{fig:observables}
\end{figure*}
%*%*%*%*%*%*%*%*%*%*%*%*%*%*%*

We compare the MST-MPS results with high-precision reference values 
obtained via the purification method \cite{zwolak2004, verstraete2004, feiguin2005}, which 
we refer to as numerically ``exact.'' 
In Figs.~\ref{fig:observables}(a-d) 
we plot the Markov entropy $S_M$ as a function of $E/N$, and $T$ dependent 
Markov free energy $F_M$, energy $E$, and magnetization $M_x$, $M_z$, which are all local quantities. 
The accuracy of the MST-MPS systematically improves with $m$ 
and for $m=4$, the data over the whole temperature range converge to the exact one. 
As we find in $S_M$, even in the vicinity of the quantum critical point at $g=1$ and $T \rightarrow 0$, 
we find a remarkable agreement (for other data at $g=1$ see Supplementary information \cite{supple}). 

Figure~\ref{fig:observables}(d) shows
the specific heat computed using the two definitions: 
\begin{gather}
    C_1 = \frac{\langle H^2 \rangle - \langle H \rangle^2}{T^2}, \label{eq:C1} \\
    C_2 = \frac{d \langle H \rangle}{dT}
    \simeq\frac{\langle H \rangle_{T+\Delta T} - \langle H \rangle_T}{\Delta T}.
    \label{eq:C2}
\end{gather}
Notice that since the MST state can accurately reproduce only the local properties of the Gibbs state, 
$C_1$ becomes less reliable, whereas $C_2$ gives a much better agreement with the exact solutions. 
The latter precision can be further improved by controlling $\Delta T$.

\subsection{How typical?}

%*%*%*%*%*%*%*%*%*%*%*%*%*%*%*
\begin{figure}
    \centering
    \includegraphics[width=1.0\hsize]{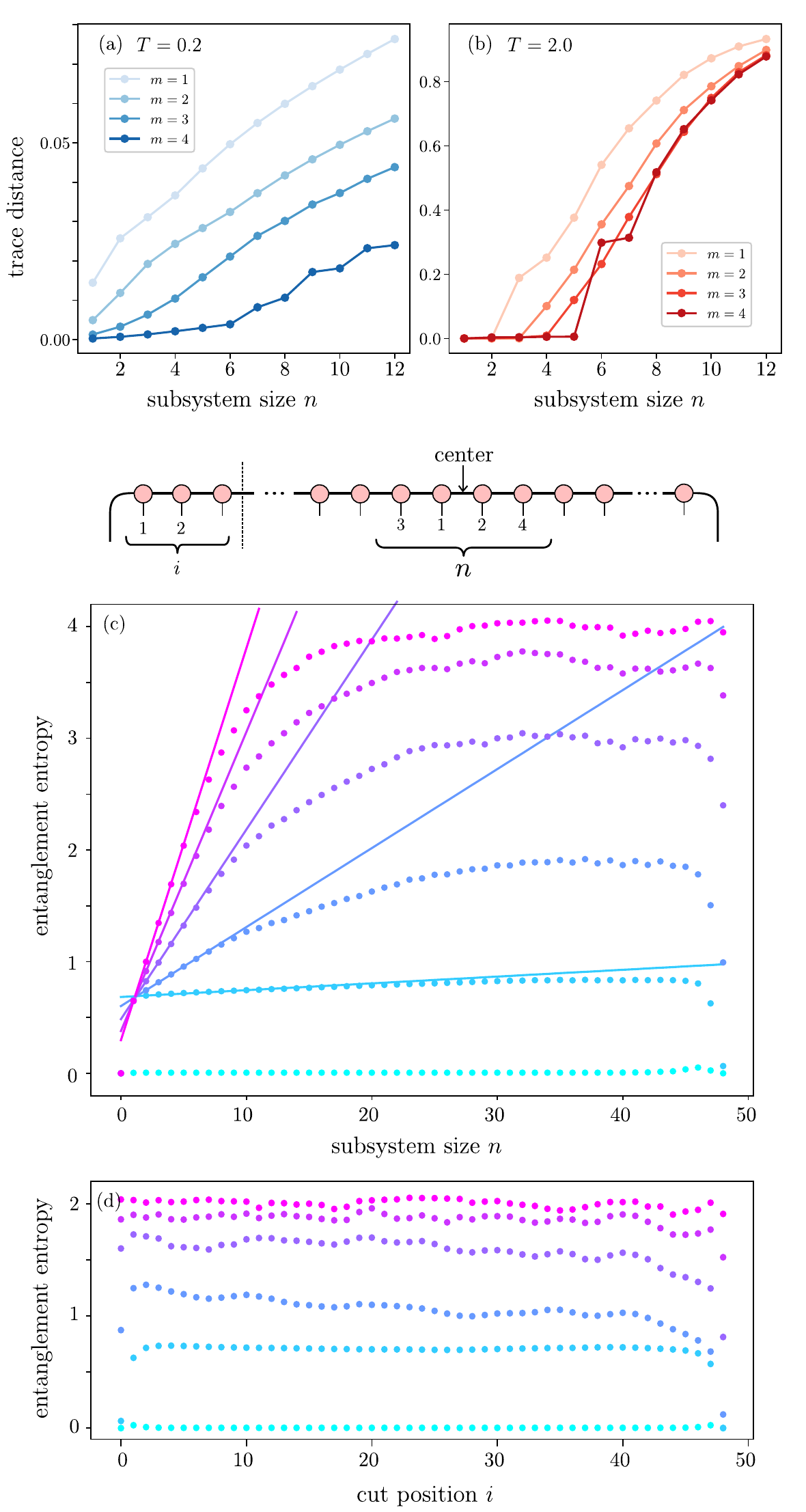}
    \caption{ 
    Transverse-field Ising chain of size $N=48$ and $g=0.5$.
    (a,b) Trace distance between local reduced density operators of the MST-MPS 
    and those of the Gibbs state for various Markov shield sizes $m = 1, 2, 3, 4$. 
    The local density operators are computed for $n$-site subsystems extracted 
    from the center of the chain at $T = 0.2$ and $2.0$. 
    Entanglement properties of the MST-MPS for various temperatures $T = 0.0$-$1.0$.
    (c) Entanglement entropy of $n$-site subsystems centered in the middle of the chain, 
    and (d) the one across each cut position $i$ along the chain. 
    }
    \label{fig:entanglement}
\end{figure}
%*%*%*%*%*%*%*%*%*%*%*%*%*%*%*

To measure the degree of canonical typicality of the MST, 
we now compute the trace distance $D(\rho, \sigma)$ 
of the reduced density operators $\rho, \sigma$ of the 
MST-MPS and the Gibbs state obtained via the purification method, respectively. 
$D(\rho, \sigma)$ gives an upper bound on the difference 
in expectation values of arbitrary observables.
In Fig.~\ref{fig:entanglement}(a), we show the trace distance at $T = 0.2$ and $g=0.5$ 
for the $n$ contiguous subsystem sites taken from the center of the system. 
The value of $D$ is less than 0.06 for most cases, 
and with increasing the size of the shield $m$ up to 4, 
$D$ is almost fully suppressed up to $n\lesssim 6$, 
demonstrating that the MST state reproduces 
the local reduced density matrices of the Gibbs state. 
Even at high enough temperature $T = 2.0$, this tendency works (see Fig.~\ref{fig:entanglement}(b)), 
while for such high $T$, the coherence of the quantum state itself is almost lost, so that 
the falling off of $D$ at $n\gtrsim m+1$ from zero does not indicate the degradation 
of the quality of the thermal state, as the physical quality in that range is determined 
by possibly up to two site correlations. 
Another observation is that $D\sim 0$ at $n \leq m+1$ for all temperature range, 
meaning that if we want to analyze the quantity supported by a certain size scale, 
we need to prepare $m$ of the same order. 
The canonical typicality of the MST is thus attained regardless of the temperature 
as far as we properly confine ourselves to the $k$-local support of the observables that rely on $D$.

\subsection{Entanglement property}

The entanglement entropy (EE) is the basic property that characterizes the quantum state. 
In Fig.~\ref{fig:entanglement}(c), we show the EE for the $n$-site subsystem from the center; 
For the first $\sim 10$ sites, the entropy grows linearly, showing the volume law entanglement, 
which saturates for larger $n$. 
This profile is almost the same as we found for the TPQ-MPS state after a random average over a few samples \cite{iwaki2021}. 
Many of the meaningful physical properties at finite temperature can be measured within the lengthscale of this volume-law range. 

Next, in Fig.~\ref{fig:entanglement}(d) we show the EE across the cut 
that divides the system into $i$ and $(N-i)$ sites on the left and right-hand sides, respectively. 
The EE profile is mostly flat and increases with temperature. 
At intermediate temperatures, there arises an asymmetry, 
which may reflect the directional nature of the Markov shielding used for MST-MPS. 
In the standard MPS without auxiliaries, EE is known to start from nearly zero 
at the edges and increase toward the center, 
known as the Page curve \cite{page1993, nakagawa2018}, 
which seriously restricted the amount of entanglement. 
Here, with the aid of auxiliaries at the edges, the almost site-independent, 
namely translationally symmetric nature is preserved, 
which we have previously found to be important for the volume law in terms of $n$ \cite{iwaki2021}. 

%%%%%%%%%%%%%%%%%%%%%%%%%%%%%%%%%%%%%%%%%%%%%%%%%%%%%%%%%%%%%%%%%
\section{Discussion}

We have established a pathway to access the most typical state within all the typical thermal pure quantum (TPQ) states 
that belong to the microcanonical shell, which is the Markov-shielded typical (MST) state. 
The TPQ states suffer inevitable random fluctuations, 
as they are generated from an infinite-temperature random state 
by cooling them down to the target temperature through the imaginary time evolution. 
These fluctuations are not numerical errors but are intrinsic and are 
significantly enhanced at lower temperatures and for smaller system sizes
\cite{iwaki2022, iwaki2024}. 
Our MST state coincides with the random averages over all these TPQ states. 
We enabled the construction of the MST state as a minimizer of the fictitious free energy 
that uses the Markov entropy, which is a local quantity, 
instead of the nonlocal von Neumann entropy. 
So far, there had been no numerical methods starting from the high-temperature side 
that smoothly connect to the ground state.  
The present scheme enables the variational approach for arbitrary temperatures, 
from the ground state to the infinite temperature, with a substantial accuracy, and more importantly, 
there is no restriction on the choice of the wave functional form 
as far as it is physical in the sense that it is globally defined consistently. 

As the previous theories of typicality are all built on randomness  
\cite{popescu2006, goldstein2006, sugita2007, reimann2007, landsman2023}, 
a typical pure state based on variational principles may appear 
incompatible with it at first glance, 
while the following interpretation is possible: 
Similarly to Landau's generalized free energies designed as functions of order parameters, 
the Markov free energy has some ``randomness'' as a phase space parameter, 
and shows a landscape with many local minima that are all regarded as the typical states. 
A set of MST states are those that continuously connect the ground state, 
except for the system that undergo a first-order transition, they continue to form the unique 
lowest free-energy manifold of states. 
In that respect, the MST  provides a deterministic pure-state representation of thermal equilibrium 
that satisfies the canonical typicality condition.

We finally make some remarks on the related works. 
Scalet \textit{et al.} \cite{scalet2025} have recently proposed a looser condition than the quantum Markov property 
to justify the MED. 
They showed that under a clustering condition, 
one can efficiently compute a state that is locally equivalent to the Gibbs state using the MED method. 
This idea can be straightforwardly applied to extend our MST approach to higher-dimensional systems. 

There have also been numerical advances very recently. 
One is the imaginary-time-evolved entangled antipodal pair (EAP) state \cite{chiba2024},  
that approaches again from the high temperature, while avoiding explicit initial randomness. 
While, this EAP state shall be an analogue of a purification method rather than the thermal pure state, 
since it is locally equivalent to a purified mixed state, 
particularly from the perspective of numerical algorithms \cite{yoneta2024}.
Another route starting from the ground state to a finite temperature was proposed 
by Cocchiarella \textit{et al.} 
using tree tensor networks \cite{cocchiarella2025}. 
The method succeeded in reaching the lowest temperature part, yielding an approximate full Gibbs state, 
while the description seemingly falls off from the thermal state at $T\gtrsim 0.1$, 
possibly because they rely on the assumption that the ground state is well described by a conformal field theory, 
or the shape of the tensor and the bond dimension may restrict the inclusion of the major excited states at higher temperatures. 
Additionally, the number of excited states that can be represented is restricted by the bond dimension of the tensor network, 
which might somehow be avoided by combining with our MST framework. 

The problem of finding an MST can be viewed from a broader perspective. 
Given a Gibbs state at some arbitrary temperature, one may rephrase our MST as 
whether there exists a pure state whose local reduced density matrices reproduce 
all $(m+1)$-local marginals of the Gibbs state. 
This question is closely related to the quantum marginal problem, 
which is known to be QMA-complete in general \cite{liu2007}. 
There, one asks whether a given collection of local density matrices is compatible with a global quantum state. 
Although several analytical approaches have been proposed \cite{yu2021}, 
the practical construction of large-scale many-body quantum states satisfying prescribed 
local constraints remains highly challenging. 
In this context, MST provides a concrete pure-state variational framework as part of the quantum marginal problem. 
We also note that certain classes of quantum marginal problems obeying a volume-law structure 
can be solved analytically by mixed states that maximize the entropy \cite{kim2021}, 
which is complementary to our approach. 
\par
Our results also suggest a promising route to representing thermal equilibrium 
on quantum devices using MITE states, which can serve as a complement to the Gibbs ensemble.
As demonstrated in Sec.~\ref{sec:numerical}, 
MST-MPS can accurately reproduce local thermal properties with relatively small bond dimensions. 
Recent advances have shown that MPS can be prepared using logarithmic-depth quantum circuits
\cite{wei2023, malz2024, scheer2025, murota2026}, 
or even constant-depth circuits with measurements and feedback 
\cite{smith2024, zhang2024, sahay2025, stephen2025, sahay2024, li2026}. 
These developments highlight the potential of MST-MPS for efficient quantum simulation of thermal properties.

\begin{acknowledgments}
A. I. thanks Akihiro Hokkyo for helpful discussions on the definition of MITE. 
A. I. was supported by the Center of Innovations for Sustainable Quantum AI
(JST Grant No. JPMJPF2221) 
and other JSPS KAKENHI (No. 25K17308).
C. H. was supported by a Grant-in-Aid for Transformative Research Areas 
``The Natural Laws of Extreme Universe---A New Paradigm for Spacetime and Matter from Quantum Information" (Grant No. 21H05191)
and other JSPS KAKENHI (No. 21K03440).
\end{acknowledgments}

\section*{Declarations}

\subsection*{Data availability}

The code used to generate the numerical results reported in this work
is publicly available on GitHub \cite{MSTcode}.

%%%%%%%%%%%%%%%%%%%%%%%%%%%%%%%%%%%%%%%%%%%%
%%%%%%%%%%%%%% bibliography %%%%%%%%%%%%%%%%
%%%%%%%%%%%%%%%%%%%%%%%%%%%%%%%%%%%%%%%%%%%%
\nocite{apsrev42Control}
\bibliographystyle{apsrev4-2}
\bibliography{references}

@CONTROL{apsrev42Control, editor="1", pages="0", title="yes", year="1"}

@article{goldstein2015,
  title = {Thermal Equilibrium of a Macroscopic Quantum System in a Pure State},
  author = {Goldstein, Sheldon and Huse, David A. and Lebowitz, Joel L. and Tumulka, Roderich},
  journal = {Phys. Rev. Lett.},
  volume = {115},
  issue = {10},
  pages = {100402},
  numpages = {5},
  year = {2015},
  month = {Sep},
  publisher = {American Physical Society},
  doi = {10.1103/PhysRevLett.115.100402},
  url = {https://link.aps.org/doi/10.1103/PhysRevLett.115.100402}
}

@article{zwolak2004,
  title = {Mixed-State Dynamics in One-Dimensional Quantum Lattice Systems: A Time-Dependent Superoperator Renormalization Algorithm},
  volume = {93},
  pages = {207205},
  ISSN = {1079-7114},
  url = {http://dx.doi.org/10.1103/PhysRevLett.93.207205},
  DOI = {10.1103/physrevlett.93.207205},
  number = {20},
  journal = {Physical Review Letters},
  publisher = {American Physical Society (APS)},
  author = {Zwolak,  Michael and Vidal,  Guifré},
  year = {2004},
  month = nov 
}

@article{verstraete2004,
  title = {Matrix Product Density Operators: Simulation of Finite-Temperature and Dissipative Systems},
  author = {Verstraete, F. and Garc\'{\i}a-Ripoll, J. J. and Cirac, J. I.},
  journal = {Phys. Rev. Lett.},
  volume = {93},
  issue = {20},
  pages = {207204},
  numpages = {4},
  year = {2004},
  month = {Nov},
  publisher = {American Physical Society},
  doi = {10.1103/PhysRevLett.93.207204},
  url = {https://link.aps.org/doi/10.1103/PhysRevLett.93.207204}
}

@misc{supple,
  note = {See Supplementary Information for details.}
}

@article{popescu2006,
  title = {Entanglement and the foundations of statistical mechanics},
  author = {Popescu, Sandu and J. Short, Anthony and Winter, Andreas},
  journal = {Nature Phys.},
  volume = {2},
  pages = {754--758},
  year = {2006},
  month = {Oct},
  doi = {10.1038/nphys444},
  url = {https://doi.org/10.1038/nphys444}
}

@article{goldstein2006,
  title = {Canonical Typicality},
  author = {Goldstein, Sheldon and Lebowitz, Joel L. and Tumulka, Roderich and Zangh\`{\i}, Nino},
  journal = {Phys. Rev. Lett.},
  volume = {96},
  issue = {5},
  pages = {050403},
  numpages = {3},
  year = {2006},
  month = {Feb},
  publisher = {American Physical Society},
  doi = {10.1103/PhysRevLett.96.050403},
  url = {https://link.aps.org/doi/10.1103/PhysRevLett.96.050403}
}

@article{sugita2007,
  title = {},
  author = {Sugita, Ayumu},
  journal = {Nonlinear Phenom. Complex Syst},
  volume = {10},
  pages = {192-195},
  year = {2007}
}

@article{reimann2007,
  title = {Typicality for Generalized Microcanonical Ensembles},
  author = {Reimann, Peter},
  journal = {Phys. Rev. Lett.},
  volume = {99},
  issue = {16},
  pages = {160404},
  numpages = {4},
  year = {2007},
  month = {Oct},
  publisher = {American Physical Society},
  doi = {10.1103/PhysRevLett.99.160404},
  url = {https://link.aps.org/doi/10.1103/PhysRevLett.99.160404}
}

@article{imada1986,
  author = {Imada, Masatoshi and Takahashi, Minoru},
  title = {Quantum Transfer Monte Carlo Method for Finite Temperature Properties and Quantum Molecular Dynamics Method for Dynamical Correlation Functions},
  journal = {Journal of the Physical Society of Japan},
  volume = {55},
  number = {10},
  pages = {3354-3361},
  year = {1986},
  doi = {10.1143/JPSJ.55.3354},
  URL = {https://doi.org/10.1143/JPSJ.55.3354}
}

@article{jaklic1994,
  title = {Lanczos method for the calculation of finite-temperature quantities in correlated systems},
  author = {Jakli\ifmmode \check{c}\else \v{c}\fi{}, J. and Prelov\ifmmode \check{s}\else \v{s}\fi{}ek, P.},
  journal = {Phys. Rev. B},
  volume = {49},
  issue = {7},
  pages = {5065--5068},
  numpages = {0},
  year = {1994},
  month = {Feb},
  publisher = {American Physical Society},
  doi = {10.1103/PhysRevB.49.5065},
  url = {https://link.aps.org/doi/10.1103/PhysRevB.49.5065}
}

@article{hams2000,
  title = {Fast algorithm for finding the eigenvalue distribution of very large matrices},
  author = {Hams, Anthony and De Raedt, Hans},
  journal = {Phys. Rev. E},
  volume = {62},
  issue = {3},
  pages = {4365--4377},
  numpages = {0},
  year = {2000},
  month = {Sep},
  publisher = {American Physical Society},
  doi = {10.1103/PhysRevE.62.4365},
  url = {https://link.aps.org/doi/10.1103/PhysRevE.62.4365}
}

@article{sugiura2012,
  title = {Thermal Pure Quantum States at Finite Temperature},
  author = {Sugiura, Sho and Shimizu, Akira},
  journal = {Phys. Rev. Lett.},
  volume = {108},
  issue = {24},
  pages = {240401},
  numpages = {4},
  year = {2012},
  month = {Jun},
  publisher = {American Physical Society},
  doi = {10.1103/PhysRevLett.108.240401},
  url = {https://link.aps.org/doi/10.1103/PhysRevLett.108.240401}
}

@article{sugiura2013,
  title = {Canonical Thermal Pure Quantum State},
  author = {Sugiura, Sho and Shimizu, Akira},
  journal = {Phys. Rev. Lett.},
  volume = {111},
  issue = {1},
  pages = {010401},
  numpages = {5},
  year = {2013},
  month = {Jul},
  publisher = {American Physical Society},
  doi = {10.1103/PhysRevLett.111.010401},
  url = {https://link.aps.org/doi/10.1103/PhysRevLett.111.010401}
}

@article{hyuga2014,
  title = {Thermal pure quantum states of many-particle systems},
  author = {Hyuga, Masahiko and Sugiura, Sho and Sakai, Kazumitsu and Shimizu, Akira},
  journal = {Phys. Rev. B},
  volume = {90},
  issue = {12},
  pages = {121110(R)},
  numpages = {5},
  year = {2014},
  month = {Sep},
  publisher = {American Physical Society},
  doi = {10.1103/PhysRevB.90.121110},
  url = {https://link.aps.org/doi/10.1103/PhysRevB.90.121110}
}

@article{poulin2011,
  title = {Markov Entropy Decomposition: A Variational Dual for Quantum Belief Propagation},
  author = {Poulin, David and Hastings, Matthew B.},
  journal = {Phys. Rev. Lett.},
  volume = {106},
  issue = {8},
  pages = {080403},
  numpages = {4},
  year = {2011},
  month = {Feb},
  publisher = {American Physical Society},
  doi = {10.1103/PhysRevLett.106.080403},
  url = {https://link.aps.org/doi/10.1103/PhysRevLett.106.080403}
}

@article{ferris2013,
  title = {Algorithms for the Markov entropy decomposition},
  author = {Ferris, Andrew J. and Poulin, David},
  journal = {Phys. Rev. B},
  volume = {87},
  issue = {20},
  pages = {205126},
  numpages = {12},
  year = {2013},
  month = {May},
  publisher = {American Physical Society},
  doi = {10.1103/PhysRevB.87.205126},
  url = {https://link.aps.org/doi/10.1103/PhysRevB.87.205126}
}

@article{scalet2025,
  title = {Classical Estimation of the Free Energy and Quantum Gibbs Sampling from the Markov Entropy Decomposition},
  author = {Scalet,  Samuel O. and Capel,  Angela and Chowdhury,  Anirban N. and Fawzi,  Hamza and Fawzi,  Omar and Kim,  Isaac H. and Tikku,  Arkin},
  year={2025},
  journal = {arXiv:2504.17405},
  url = {https://arxiv.org/abs/2504.17405}
}

@article{kato2019,
  title = {Quantum Approximate Markov Chains are Thermal},
  author = {Kato, Kohtaro and Brand\~ao, Fernando G. S. L.},
  journal = {Communications in Mathematical Physics},
  volume = {370},
  issue = {1},
  pages = {117-149},
  numpages = {32},
  year = {2019},
  month = {Aug},
  doi = {10.1007/s00220-019-03485-6},
  url = {https://doi.org/10.1007/s00220-019-03485-6}
}

@article{kuwahara2025,
  title = {Clustering of Conditional Mutual Information and Quantum Markov Structure at Arbitrary Temperatures},
  volume = {15},
  pages = {041010},
  ISSN = {2160-3308},
  url = {http://dx.doi.org/10.1103/9hx7-pzxw},
  DOI = {10.1103/9hx7-pzxw},
  number = {4},
  journal = {Physical Review X},
  publisher = {American Physical Society (APS)},
  author = {Kuwahara,  Tomotaka},
  year = {2025},
  month = oct 
}

@article{kato2025,
  title={On the Clustering of Conditional Mutual Information via Dissipative Dynamics},
  author={Kato, Kohtaro and Kuwahara, Tomotaka},
  year={2025},
  journal = {arXiv:2504.02235},
  url = {https://arxiv.org/abs/2504.02235}
}

@article{chen2025,
  title={Quantum Gibbs states are locally Markovian},
  author={Chen, Chi-Fang and Rouz\'{e}, Cambyse},
  year={2025},
  journal = {arXiv:2504.02208},
  url = {https://arxiv.org/abs/2504.02208}
}

@article{yang2026,
  title={Improved estimate of local Markovianity for quantum Gibbs states},
  author={Tai Hsuan Yang},
  year={2025},
  journal = {arXiv.2609.38007},
  url = {https://arxiv.org/abs/2609.38007}
}

@article{white1992,
  title = {Density matrix formulation for quantum renormalization groups},
  author = {White, Steven R.},
  journal = {Phys. Rev. Lett.},
  volume = {69},
  issue = {19},
  pages = {2863--2866},
  numpages = {0},
  year = {1992},
  month = {Nov},
  publisher = {American Physical Society},
  doi = {10.1103/PhysRevLett.69.2863},
  url = {https://link.aps.org/doi/10.1103/PhysRevLett.69.2863}
}

@article{white1993,
  title = {Density-matrix algorithms for quantum renormalization groups},
  author = {White, Steven R.},
  journal = {Phys. Rev. B},
  volume = {48},
  issue = {14},
  pages = {10345--10356},
  numpages = {0},
  year = {1993},
  month = {Oct},
  publisher = {American Physical Society},
  doi = {10.1103/PhysRevB.48.10345},
  url = {https://link.aps.org/doi/10.1103/PhysRevB.48.10345}
}

@article{fannes1992,
  title = {Finitely correlated states on quantum spin chains},
  author = {M. Fannes and B. Nachtergaele and R. F. Werner},
  journal = {Commun. Math. Phys.},
  volume = {144},
  issue = {3},
  pages = {443-490},
  year = {1992},
  month = {Mar},
  doi = {10.1007/BF02099178},
  url = {https://doi.org/10.1007/BF02099178}
}

@article{cocchiarella2025,
  title = {Low-temperature Gibbs states with tensor networks},
  volume = {7},
  pages = {033184},
  ISSN = {2643-1564},
  url = {http://dx.doi.org/10.1103/x2gx-wsp1},
  DOI = {10.1103/x2gx-wsp1},
  number = {3},
  journal = {Physical Review Research},
  publisher = {American Physical Society (APS)},
  author = {Cocchiarella,  Denise and Bañuls,  Mari Carmen},
  year = {2025},
  month = aug 
}

@book{liu2006,
  title={Proc. RANDOM},
  author={Liu, Y.-K.},
  year={2006},
  publisher={Springer-Verlag, Berlin}, 
  pages = {438},
}

@inproceedings{schilling2014,
  title = {The quantum marginal problem},
  url = {http://dx.doi.org/10.1142/9789814618144_0010},
  DOI = {10.1142/9789814618144_0010},
  booktitle = {Mathematical Results in Quantum Mechanics},
  publisher = {WORLD SCIENTIFIC},
  author = {Schilling,  C.},
  year = {2014},
  month = nov,
  pages = {165–176}
}

@article{liu2007,
  title = {Quantum Computational Complexity of the $N$-Representability Problem: QMA Complete},
  author = {Liu, Yi-Kai and Christandl, Matthias and Verstraete, F.},
  journal = {Phys. Rev. Lett.},
  volume = {98},
  issue = {11},
  pages = {110503},
  numpages = {4},
  year = {2007},
  month = {Mar},
  publisher = {American Physical Society},
  doi = {10.1103/PhysRevLett.98.110503},
  url = {https://link.aps.org/doi/10.1103/PhysRevLett.98.110503}
}

@article{yu2021,
  title = {A complete hierarchy for the pure state marginal problem in quantum mechanics},
  volume = {12},
  pages = {1012},
  ISSN = {2041-1723},
  url = {http://dx.doi.org/10.1038/s41467-020-20799-5},
  DOI = {10.1038/s41467-020-20799-5},
  number = {1},
  journal = {Nature Communications},
  publisher = {Springer Science and Business Media LLC},
  author = {Yu,  Xiao-Dong and Simnacher,  Timo and Wyderka,  Nikolai and Nguyen,  H. Chau and G\"{u}hne,  Otfried},
  year = {2021},
  month = Feb 
}

@article{kim2021,
  title = {Entropy Scaling Law and the Quantum Marginal Problem},
  volume = {11},
  pages = {021039},
  ISSN = {2160-3308},
  url = {http://dx.doi.org/10.1103/PhysRevX.11.021039},
  DOI = {10.1103/physrevx.11.021039},
  number = {2},
  journal = {Physical Review X},
  publisher = {American Physical Society (APS)},
  author = {Kim,  Isaac H.},
  year = {2021},
  month = May 
}

@article{iwaki2022,
  title = {Purity of thermal mixed quantum states},
  author = {Iwaki, Atsushi and Hotta, Chisa},
  journal = {Phys. Rev. B},
  volume = {106},
  issue = {9},
  pages = {094409},
  numpages = {18},
  year = {2022},
  month = {Sep},
  publisher = {American Physical Society},
  doi = {10.1103/PhysRevB.106.094409},
  url = {https://link.aps.org/doi/10.1103/PhysRevB.106.094409}
}

@article{iwaki2024,
  title = {Sample complexity of matrix product states at finite temperature},
  author = {Iwaki, Atsushi and Hotta, Chisa},
  journal = {Phys. Rev. B},
  volume = {109},
  issue = {22},
  pages = {224410},
  numpages = {11},
  year = {2024},
  month = {Jun},
  publisher = {American Physical Society},
  doi = {10.1103/PhysRevB.109.224410},
  url = {https://link.aps.org/doi/10.1103/PhysRevB.109.224410}
}

@article{chiba2024,
  title = {Exact Thermal Eigenstates of Nonintegrable Spin Chains at Infinite Temperature},
  author = {Chiba, Yuuya and Yoneta, Yasushi},
  journal = {Phys. Rev. Lett.},
  volume = {133},
  issue = {17},
  pages = {170404},
  numpages = {6},
  year = {2024},
  month = {Oct},
  publisher = {American Physical Society},
  doi = {10.1103/PhysRevLett.133.170404},
  url = {https://link.aps.org/doi/10.1103/PhysRevLett.133.170404}
}

@article{yoneta2024,
  title = {Thermal pure states for systems with antiunitary symmetries and their tensor network representations},
  author = {Yoneta, Yasushi},
  journal = {Phys. Rev. Res.},
  volume = {6},
  issue = {4},
  pages = {L042062},
  numpages = {7},
  year = {2024},
  month = {Dec},
  publisher = {American Physical Society},
  doi = {10.1103/PhysRevResearch.6.L042062},
  url = {https://link.aps.org/doi/10.1103/PhysRevResearch.6.L042062}
}

@article{doroudiani2020,
  title = {Planar maximally entangled states},
  author = {Doroudiani, Mehregan and Karimipour, Vahid},
  journal = {Phys. Rev. A},
  volume = {102},
  issue = {1},
  pages = {012427},
  numpages = {8},
  year = {2020},
  month = {Jul},
  publisher = {American Physical Society},
  doi = {10.1103/PhysRevA.102.012427},
  url = {https://link.aps.org/doi/10.1103/PhysRevA.102.012427}
}

@article{wang2021,
  title = {Planar k-uniform states: a generalization of planar maximally entangled states},
  volume = {20},
  ISSN = {1573-1332},
  url = {http://dx.doi.org/10.1007/s11128-021-03204-y},
  DOI = {10.1007/s11128-021-03204-y},
  number = {8},
  journal = {Quantum Information Processing},
  publisher = {Springer Science and Business Media LLC},
  author = {Wang,  Yan-Ling},
  year = {2021},
  month = Aug,
  pages = {271}
}

@article{mohapatra2025,
  title = {Exact Volume-Law Entangled Zero-Energy Eigenstates in a Large Class of Spin Models},
  author = {Mohapatra, Sashikanta and Moudgalya, Sanjay and Balram, Ajit C.},
  journal = {Phys. Rev. Lett.},
  volume = {134},
  issue = {21},
  pages = {210403},
  numpages = {7},
  year = {2025},
  month = {May},
  publisher = {American Physical Society},
  doi = {10.1103/PhysRevLett.134.210403},
  url = {https://link.aps.org/doi/10.1103/PhysRevLett.134.210403}
}

@article{hokkyo2026,
  title = {Exact Thermal Stabilizer Eigenstates at Infinite Temperature},
  author = {Hokkyo, Akihiro},
  year = {2026},
  journal = {arXiv:2601.16177},
  url = {https://arxiv.org/abs/2601.16177}
}

@unpublished{hammersley1971,
  author = {Hammersley, J. M. and Clifford, P.},
  title = {Markov Fields on Finite Graphs and Lattices},
  year = {1971},
  note = {Unpublished manuscript},
  url = {https://ora.ox.ac.uk/objects/uuid:4ea849da-1511-4578-bb88-6a8d02f457a6}
}

@article{brown2012,
  title = {Quantum Markov Networks and Commuting Hamiltonians},
  author = {Brown, Winton and Poulin, David},
  year = {2012},
  journal = {arXiv:1206.0755},
  url = {https://arxiv.org/abs/1206.0755}
}

@article{carleo2017,
  title = {Solving the quantum many-body problem with artificial neural networks},
  volume = {355},
  ISSN = {1095-9203},
  url = {http://dx.doi.org/10.1126/science.aag2302},
  DOI = {10.1126/science.aag2302},
  number = {6325},
  journal = {Science},
  publisher = {American Association for the Advancement of Science (AAAS)},
  author = {Carleo,  Giuseppe and Troyer,  Matthias},
  year = {2017},
  month = feb,
  pages = {602–606}
}

@book{kitaev2002,
  title={Classical and Quantum Computation},
  author={Kitaev, A.Y. and Shen, A. and Vyalyi, M.N.},
  isbn={9780821832295},
  lccn={2002016686},
  series={Graduate studies in mathematics},
  url={https://books.google.co.jp/books?id=08vZYhafYEAC},
  year={2002},
  publisher={American Mathematical Society}
}

@article{aharonov2009,
  title = {The Power of Quantum Systems on a Line},
  author = {Aharonov, Dorit and Gottesman, Daniel and Irani, Sandy and Kempe, Julia},
  journal = {Communications in Mathematical Physics},
  volume = {287},
  issue = {1},
  pages = {41--65},
  year = {2009},
  doi = {10.1007/s00220-008-0710-3},
  url = {https://doi.org/10.1007/s00220-008-0710-3}
}

@article{hallgren2013,
author = {Hallgren, Sean and Nagaj, Daniel and Narayanaswami, Sandeep},
title = {The Local Hamiltonian Problem on a Line with Eight States is QMA-Complete},
year = {2013},
issue_date = {September 2013},
publisher = {Rinton Press, Incorporated},
address = {Paramus, NJ},
volume = {13},
number = {9–10},
issn = {1533-7146},
journal = {Quantum Info. Comput.},
month = {sep},
pages = {721–750},
numpages = {30}
}

@article{lieb1973,
  title = {Proof of the strong subadditivity of quantum-mechanical entropy},
  volume = {14},
  ISSN = {1089-7658},
  url = {http://dx.doi.org/10.1063/1.1666274},
  DOI = {10.1063/1.1666274},
  number = {12},
  journal = {Journal of Mathematical Physics},
  publisher = {AIP Publishing},
  author = {Lieb,  Elliott H. and Ruskai,  Mary Beth},
  year = {1973},
  month = dec,
  pages = {1938–1941}
}

@article{feiguin2005,
  title = {Finite-temperature density matrix renormalization using an enlarged Hilbert space},
  author = {Feiguin, Adrian E. and White, Steven R.},
  journal = {Phys. Rev. B},
  volume = {72},
  issue = {22},
  pages = {220401},
  numpages = {4},
  year = {2005},
  month = {Dec},
  publisher = {American Physical Society},
  doi = {10.1103/PhysRevB.72.220401},
  url = {https://link.aps.org/doi/10.1103/PhysRevB.72.220401}
}

@Article{itensor1,
	title={{The ITensor Software Library for Tensor Network Calculations}},
	author={Matthew Fishman and Steven R. White and E. Miles Stoudenmire},
	journal={SciPost Phys. Codebases},
	pages={4},
	year={2022},
	publisher={SciPost},
	doi={10.21468/SciPostPhysCodeb.4},
	url={https://scipost.org/10.21468/SciPostPhysCodeb.4},
}

@Article{itensor2,
	title={{Codebase release 0.3 for ITensor}},
	author={Matthew Fishman and Steven R. White and E. Miles Stoudenmire},
	journal={SciPost Phys. Codebases},
	pages={4-r0.3},
	year={2022},
	publisher={SciPost},
	doi={10.21468/SciPostPhysCodeb.4-r0.3},
	url={https://scipost.org/10.21468/SciPostPhysCodeb.4-r0.3},
}

@article{page1993,
  title = {Average entropy of a subsystem},
  author = {Page, Don N.},
  journal = {Phys. Rev. Lett.},
  volume = {71},
  issue = {9},
  pages = {1291--1294},
  numpages = {0},
  year = {1993},
  month = {Aug},
  publisher = {American Physical Society},
  doi = {10.1103/PhysRevLett.71.1291},
  url = {https://link.aps.org/doi/10.1103/PhysRevLett.71.1291}
}

@article{nakagawa2018,
  title = {Universality in volume-law entanglement of scrambled pure quantum states},
  author = {Nakagawa, Yuya O. and Watanabe, Masataka and Fujita, Hiroyuki and Sugiura, Sho},
  journal = {Nature Communications},
  volume = {9},
  issue = {1},
  pages = {1635},
  year = {2018},
  month = {Apr},
  doi = {10.1038/s41467-018-03883-9},
  url = {https://doi.org/10.1038/s41467-018-03883-9}
}

@article{landsman2023,
  title = {Typical = Random},
  volume = {12},
  ISSN = {2075-1680},
  url = {http://dx.doi.org/10.3390/axioms12080727},
  DOI = {10.3390/axioms12080727},
  number = {8},
  journal = {Axioms},
  publisher = {MDPI AG},
  author = {Landsman,  Klaas},
  year = {2023},
  month = jul,
  pages = {727}
}

@article{deutsch1991,
  title = {Quantum statistical mechanics in a closed system},
  author = {Deutsch, J. M.},
  journal = {Phys. Rev. A},
  volume = {43},
  issue = {4},
  pages = {2046--2049},
  numpages = {0},
  year = {1991},
  month = {Feb},
  publisher = {American Physical Society},
  doi = {10.1103/PhysRevA.43.2046},
  url = {https://link.aps.org/doi/10.1103/PhysRevA.43.2046}
}

@article{srednicki1994,
  title = {Chaos and quantum thermalization},
  author = {Srednicki, Mark},
  journal = {Phys. Rev. E},
  volume = {50},
  issue = {2},
  pages = {888--901},
  numpages = {0},
  year = {1994},
  month = {Aug},
  publisher = {American Physical Society},
  doi = {10.1103/PhysRevE.50.888},
  url = {https://link.aps.org/doi/10.1103/PhysRevE.50.888}
}

@article{rigol2008,
  title = {Thermalization and its mechanism for generic isolated quantum systems},
  volume = {452},
  ISSN = {1476-4687},
  url = {http://dx.doi.org/10.1038/nature06838},
  DOI = {10.1038/nature06838},
  number = {7189},
  journal = {Nature},
  publisher = {Springer Science and Business Media LLC},
  author = {Rigol,  Marcos and Dunjko,  Vanja and Olshanii,  Maxim},
  year = {2008},
  month = apr,
  pages = {854–858}
}

@article{petz1986,
  title = {Sufficient subalgebras and the relative entropy of states of a von Neumann algebra},
  volume = {105},
  ISSN = {1432-0916},
  url = {http://dx.doi.org/10.1007/BF01212345},
  DOI = {10.1007/bf01212345},
  number = {1},
  journal = {Communications in Mathematical Physics},
  publisher = {Springer Science and Business Media LLC},
  author = {Petz,  D\'{e}nes},
  year = {1986},
  month = mar,
  pages = {123–131}
}

@article{petz1988,
  title = {SUFFICIENCY OF CHANNELS OVER VON NEUMANN ALGEBRAS},
  volume = {39},
  ISSN = {1464-3847},
  url = {http://dx.doi.org/10.1093/qmath/39.1.97},
  DOI = {10.1093/qmath/39.1.97},
  number = {1},
  journal = {The Quarterly Journal of Mathematics},
  publisher = {Oxford University Press (OUP)},
  author = {Petz,  D\'{e}nes},
  year = {1988},
  pages = {97–108}
}

@article{mcmillan1965,
  title = {Ground State of Liquid ${\mathrm{He}}^{4}$},
  author = {McMillan, W. L.},
  journal = {Phys. Rev.},
  volume = {138},
  issue = {2A},
  pages = {A442--A451},
  numpages = {0},
  year = {1965},
  month = {Apr},
  publisher = {American Physical Society},
  doi = {10.1103/PhysRev.138.A442},
  url = {https://link.aps.org/doi/10.1103/PhysRev.138.A442}
}

@article{mogensen2018, 
  doi = {10.21105/joss.00615}, 
  url = {https://doi.org/10.21105/joss.00615}, 
  year = {2018}, 
  publisher = {The Open Journal}, 
  volume = {3}, 
  number = {24}, 
  pages = {615}, 
  author = {Mogensen, Patrick K. and Riseth, Asbjørn N.}, 
  title = {Optim: A mathematical optimization package for Julia}, 
  journal = {Journal of Open Source Software}
}

@article{iwaki2021,
  title = {Thermal pure quantum matrix product states recovering a volume law entanglement},
  author = {Iwaki, Atsushi and Shimizu, Akira and Hotta, Chisa},
  journal = {Phys. Rev. Res.},
  volume = {3},
  issue = {2},
  pages = {L022015},
  numpages = {6},
  year = {2021},
  month = {May},
  publisher = {American Physical Society},
  doi = {10.1103/PhysRevResearch.3.L022015},
  url = {https://link.aps.org/doi/10.1103/PhysRevResearch.3.L022015}
}

@article{chen2025-2,
  title = {Efficient quantum thermal simulation},
  volume = {646},
  ISSN = {1476-4687},
  url = {http://dx.doi.org/10.1038/s41586-025-09583-x},
  DOI = {10.1038/s41586-025-09583-x},
  number = {8085},
  journal = {Nature},
  publisher = {Springer Science and Business Media LLC},
  author = {Chen,  Chi-Fang and Kastoryano,  Michael and Brandão,  Fernando G. S. L. and Gilyén,  András},
  year = {2025},
  month = Oct,
  pages = {561–566}
}

@article{rouze2026-1,
  title = {Optimal Quantum Algorithm for Gibbs State Preparation},
  author = {Rouz\'e, Cambyse and Stilck Fran\ifmmode \mbox{\c{c}}\else \c{c}\fi{}a, Daniel and Alhambra, \'Alvaro M.},
  journal = {Phys. Rev. Lett.},
  volume = {136},
  issue = {6},
  pages = {060601},
  numpages = {6},
  year = {2026},
  month = {Feb},
  publisher = {American Physical Society},
  doi = {10.1103/lhht-svmn},
  url = {https://link.aps.org/doi/10.1103/lhht-svmn}
}

@article{rouze2026-2,
  title = {Efficient thermalization and universal quantum computing with quantum Gibbs samplers},
  ISSN = {1745-2481},
  url = {http://dx.doi.org/10.1038/s41567-026-03246-y},
  DOI = {10.1038/s41567-026-03246-y},
  journal = {Nature Physics},
  publisher = {Springer Science and Business Media LLC},
  author = {Rouz\'{e},  Cambyse and Stilck Fran\c{c}a,  Daniel and Alhambra,  {\'A}lvaro M.},
  year = {2026},
  month = Apr, 
  volume = {22},
  pages = {1123}
}

@article{winter2016,
  title = {Tight Uniform Continuity Bounds for Quantum Entropies: Conditional Entropy,  Relative Entropy Distance and Energy Constraints},
  volume = {347},
  ISSN = {1432-0916},
  url = {http://dx.doi.org/10.1007/s00220-016-2609-8},
  DOI = {10.1007/s00220-016-2609-8},
  number = {1},
  journal = {Communications in Mathematical Physics},
  publisher = {Springer Science and Business Media LLC},
  author = {Winter,  Andreas},
  year = {2016},
  month = Mar,
  pages = {291–313}
}

@article{wei2023,
  title = {Efficient adiabatic preparation of tensor network states},
  volume = {5},
  pages = {L022037},
  ISSN = {2643-1564},
  url = {http://dx.doi.org/10.1103/PhysRevResearch.5.L022037},
  DOI = {10.1103/physrevresearch.5.l022037},
  number = {2},
  journal = {Physical Review Research},
  publisher = {American Physical Society (APS)},
  author = {Wei,  Zhi-Yuan and Malz,  Daniel and Cirac,  J. Ignacio},
  year = {2023},
  month = May 
}

@article{malz2024,
  title = {Preparation of Matrix Product States with Log-Depth Quantum Circuits},
  volume = {132},
  pages = {040404},
  ISSN = {1079-7114},
  url = {http://dx.doi.org/10.1103/PhysRevLett.132.040404},
  DOI = {10.1103/physrevlett.132.040404},
  number = {4},
  journal = {Physical Review Letters},
  publisher = {American Physical Society (APS)},
  author = {Malz,  Daniel and Styliaris,  Georgios and Wei,  Zhi-Yuan and Cirac,  J. Ignacio},
  year = {2024},
  month = Jan 
}

@article{scheer2025,
  title = {Renormalization-group-based preparation of matrix product states on up to 80 qubits},
  author = {Scheer,  Moritz and Baiardi,  Alberto and Marty,  Elisa B\"{a}umer and Wei,  Zhi-Yuan and Malz,  Daniel},
  year = {2025},
  journal = {arXiv:2510.24681},
  url = {https://arxiv.org/abs/2510.24681}
}

@article{murota2026,
  title = {Exact log-depth preparation of highly entangled matrix product states},
  author = {Murota,  Keisuke and Sauvage,  Frédéric and Ballarin,  Marco and Matos,  Gabriel and Rinaldi,  Enrico},
  year = {2026},
  journal = {arXiv:2606.24475},
  url = {https://arxiv.org/abs/2606.24475}
}

@article{smith2024,
  title = {Constant-Depth Preparation of Matrix Product States with Adaptive Quantum Circuits},
  volume = {5},
  pages = {030344},
  ISSN = {2691-3399},
  url = {http://dx.doi.org/10.1103/PRXQuantum.5.030344},
  DOI = {10.1103/prxquantum.5.030344},
  number = {3},
  journal = {PRX Quantum},
  publisher = {American Physical Society (APS)},
  author = {Smith,  Kevin C. and Khan,  Abid and Clark,  Bryan K. and Girvin,  S.M. and Wei,  Tzu-Chieh},
  year = {2024},
  month = Sept 
}

@article{zhang2024,
  title = {Characterizing Matrix-Product States and Projected Entangled-Pair States Preparable via Measurement and Feedback},
  volume = {5},
  pages = {040304},
  ISSN = {2691-3399},
  url = {http://dx.doi.org/10.1103/PRXQuantum.5.040304},
  DOI = {10.1103/prxquantum.5.040304},
  number = {4},
  journal = {PRX Quantum},
  publisher = {American Physical Society (APS)},
  author = {Zhang,  Yifan and Gopalakrishnan,  Sarang and Styliaris,  Georgios},
  year = {2024},
  month = Oct 
}

@article{sahay2025,
  title = {Classifying One-Dimensional Quantum States Prepared by a Single Round of Measurements},
  volume = {6},
  pages = {010329},
  ISSN = {2691-3399},
  url = {http://dx.doi.org/10.1103/PRXQuantum.6.010329},
  DOI = {10.1103/prxquantum.6.010329},
  number = {1},
  journal = {PRX Quantum},
  publisher = {American Physical Society (APS)},
  author = {Sahay,  Rahul and Verresen,  Ruben},
  year = {2025},
  month = Feb 
}

@article{stephen2025,
  title = {Preparing matrix product states via fusion: Constraints and extensions},
  volume = {112},
  pages = {235127},
  ISSN = {2469-9969},
  url = {http://dx.doi.org/10.1103/cv3q-5l8w},
  DOI = {10.1103/cv3q-5l8w},
  number = {23},
  journal = {Physical Review B},
  publisher = {American Physical Society (APS)},
  author = {Stephen,  David T. and Hart,  Oliver},
  year = {2025},
  month = Dec 
}

@article{sahay2024,
  title = {Finite-Depth Preparation of Tensor Network States from Measurement},
  author={Rahul Sahay and Ruben Verresen},
  year = {2024},
  journal = {arXiv:2404.17087},
  url = {https://arxiv.org/abs/2404.17087}
}

@article{li2026,
  title={State preparation via measurement and feedback: pushing relations, state structures, and non-invertible symmetries}, 
  author={Yabo Li and Aditi Mitra and Tsung-Cheng Lu},
  year = {2026},
  journal = {arXiv:2608.08821},
  url = {https://arxiv.org/abs/2608.08821}
}

@misc{MSTcode,
  author = {Iwaki, Atsushi},
  title = {{Markov-shielded typicality for 1D spin chains}},
  howpublished = {GitHub repository, \url{https://github.com/AtsushiIwaki/markov-shielded-typicality}},
  year = {2026}
}

\end{document}

\title{Supplementary Information of \\ ``Markov-Shielded Typicality for Thermal Quantum States''}
\author{Atsushi Iwaki}
\email{atsushi-iwaki@phys.s.u-tokyo.ac.jp}
\author{Chisa Hotta}
\maketitle

\setcounter{section}{0}
\setcounter{equation}{0}
\setcounter{figure}{0}
\setcounter{table}{0}
\renewcommand{\thesection}{S\arabic{section}}
\renewcommand{\theequation}{S\arabic{equation}}
\renewcommand{\thefigure}{S\arabic{figure}}
\renewcommand{\thetable}{S\arabic{table}}
\renewcommand{\thesubsection}{S\arabic{subsection}}

\subsection{Optimization of MST-MPS}

In this section, we describe the implementation details of the MST-MPS optimization. 
In our implementation based on ITensors.jl \cite{itensor1, itensor2}, 
the MPS with auxiliary degrees of freedom is stored as a standard MPS defined on an enlarged physical Hilbert space. 
More specifically, 
we introduce $N_{\mathrm{aux}}=\log_2 \chi$ auxiliary sites on both the left and right sides. 
Accordingly, the total number of sites in the MPS is given by $N_{\mathrm{tot}}=N+2N_{\mathrm{aux}}$. 
For this MPS with auxiliary sites, we define a function that evaluates the Markov free-energy functional. 
We then optimize each of the $N_{\mathrm{tot}}$ tensors in sequence 
using the BFGS algorithm implemented in Optim.jl \cite{mogensen2018}, 
sweeping from the left edge to the right edge and then back again. 
One such round trip is defined as a single sweep. 
The optimization is repeated over multiple sweeps 
and terminated when the change in the Markov free energy per sites becomes smaller than $10^{-4}$.

\begin{figure}
\centering
    \includegraphics[width=\hsize]{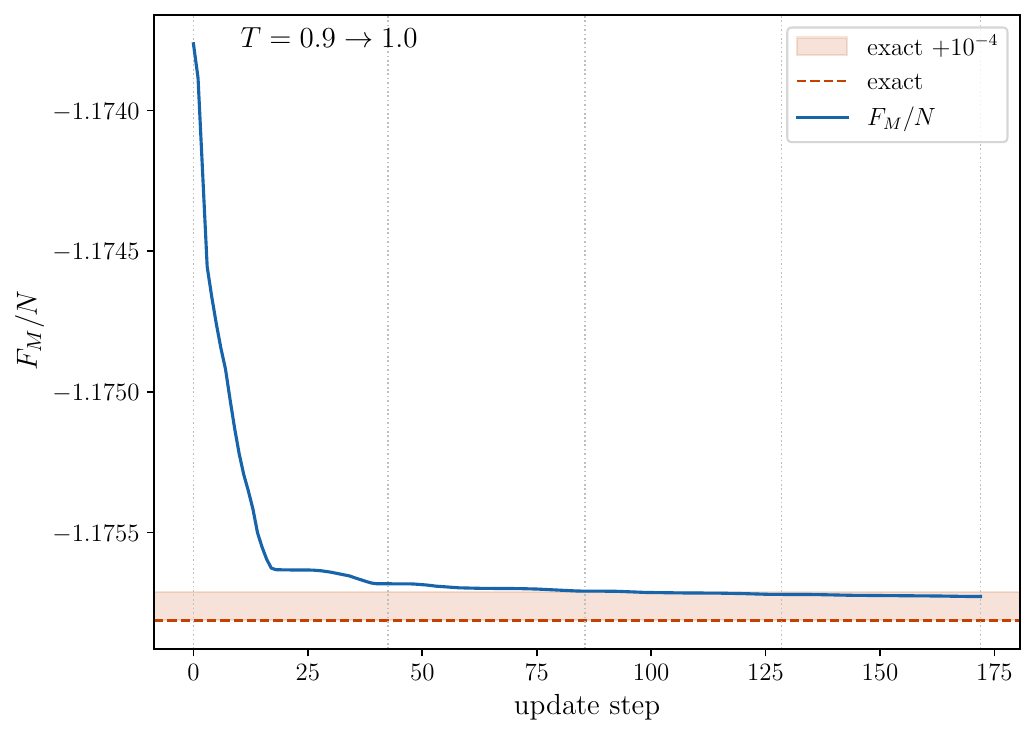}
    \caption{
    Behavior of the cost function, namely the Markov free-energy functional, 
    during the tensor updates in the MST-MPS optimization. 
    Results are shown for the one-dimensional transverse-field Ising model with $N=16$ and $g=0.5$, 
    when the temperature is raised from $T=0.9$ to $T=1.0$. 
    A single sweep consists of updating the tensors from the left edge to the right edge and back again. 
    The plot shows four sweeps in total, with the boundaries between sweeps marked by gray dotted lines. 
    The exact value obtained by the purification method is indicated by the orange dashed line, 
    while the shaded region represents values within $10^{-4}$ of the exact result.
    }
\label{fig:optimization}
\end{figure}

Figure~\ref{fig:optimization} shows the behavior of the cost function, 
namely the Markov free-energy functional, 
during each local tensor update for the case of $N=16$, $\chi=8$, and $m=3$, 
when the temperature is increased from $T=0.9$ to $T=1.0$. 
For clarity, in this example we do not terminate the optimization according to the free-energy tolerance, 
and instead explicitly show the results for 4 sweeps. 
The cost function decreases rapidly during the 1st sweep, although the accuracy is not yet sufficient at that stage. 
After 2 sweeps, the result reaches an accuracy of approximately $10^{-4}$ compared with the exact value. 
After the 3rd and 4th sweeps, the value changes only marginally and appears to be saturated. 
This residual deviation should be interpreted not as a failure of the optimization, 
but rather as the intrinsic limitation of the free-energy approximation achievable within the MST-MPS ansatz. 
For practical purposes, the numerical accuracy is already sufficient at this stage.

%*%*%*%*%*%*%*%*%*%*%*%*%*%*%*%*%*%*%*%*%*%*%*%*%*%*%*%*%*%
%*%*%*%*%*%*%*%*%*%*%*%*%*%*%*%*%*%*%*%*%*%*%*%*%*%*%*%*%*%

\subsection{MST-MPS calculations at the critical point}

To support the reliability of MST-MPS shown in Fig.~5 in the main text, 
we perform the same set of calculations for different parameters of the transverse-field Ising chain. 
The value $g=0.5$, mainly used in the main-text figures, corresponds to a gapped ground state in one dimension. 
Therefore, we show the results of the gapless critical point $g=1$, 
for which the relation between $S_M/N$ and $E/N$ is already shown in Fig.~5(a) of the main text.

\begin{figure*}
\centering
    \includegraphics[width=\hsize]{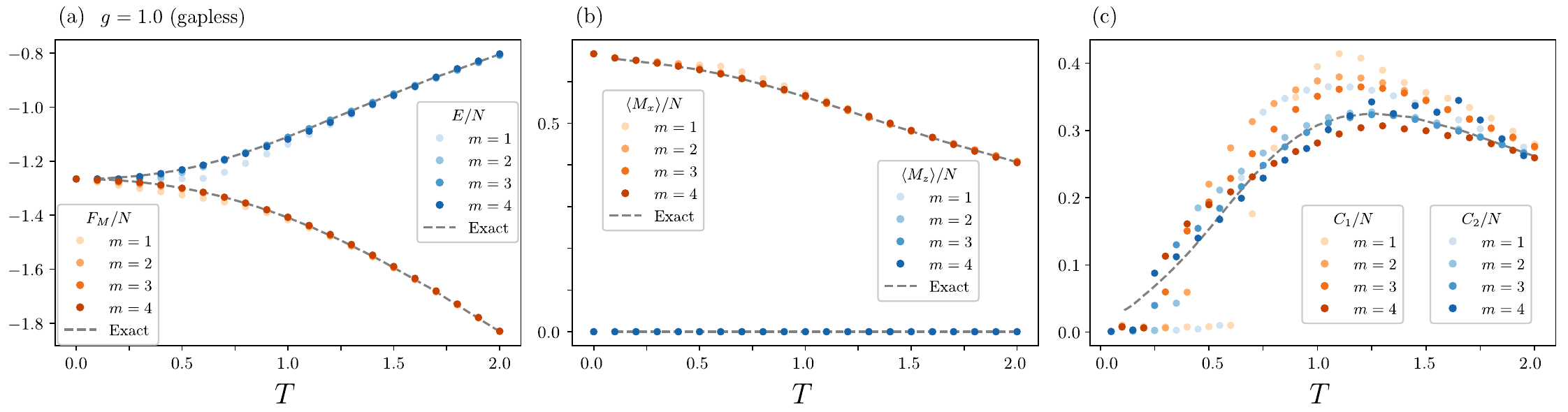}
    \caption{
    Physical quantities of the transverse-field Ising chain for $N=48$ at the critical point $g=1.0$.
    We use the MST-MPS method 
    for various sizes of Markov shields $m = 1, 2, 3, 4$ 
    and the purification method \cite{feiguin2005} 
    as the exact results (dashed lines). 
    Temperature dependence of 
    (a) the Markov free energy $F_M/N$ and the energy per site $E/N$, 
    (b) the transverse and longitudinal magnetizations, $\langle M_x \rangle/N$ and $\langle M_z \rangle/N$, 
    and (c) the specific heat per site, evaluated in two different ways: 
    $C_1$ from Hamiltonian fluctuations and $C_2$ from the temperature derivative of the energy.
    }
\label{fig:g1}
\end{figure*}

Figure~\ref{fig:g1} shows the temperature dependence of the Markov free energy $F_M / N$, 
the energy per site $E/N$, and the magnetizations $\langle M_x\rangle/N$ and $\langle M_z\rangle/N$
for $N=48$ and $m=1,2,3,4$, allowing for a direct comparison with Fig.~5(b-d) of the main text.
In contrast to what might be expected near criticality, 
the MST-MPS results do not show a substantial deterioration in accuracy 
or an increase in fluctuations compared with the $g=0.5$ case, 
although the correlation length $\xi$, 
which strongly affects the quantum Markov property, 
is expected to diverge near the critical point.

%*%*%*%*%*%*%*%*%*%*%*%*%*%*%*%*%*%*%*%*%*%*%*%*%*%*%*%*%*%
%*%*%*%*%*%*%*%*%*%*%*%*%*%*%*%*%*%*%*%*%*%*%*%*%*%*%*%*%*%

\subsection{Entanglement with auxiliary sites}

The MST-MPS method proposed in this work is initialized from the ground state. 
Accordingly, we adopt an MPS ansatz with auxiliary sites as the variational manifold. 
At zero temperature, the physical and auxiliary degrees of freedom are completely disentangled, 
so that the effective bond dimension connected to the auxiliary sector is equal to one. 
In practice, the corresponding unused part of the variational space is represented numerically by zero-filled components.
As the temperature is raised, entanglement between the physical system and the auxiliary sites is progressively generated. 
This auxiliary entanglement provides the variational freedom necessary 
to construct an MST state that is locally equivalent to the Gibbs state. 
At the same time, the effective bond dimension in the auxiliary sector becomes larger as the temperature increases.

\begin{figure}
\centering
    \includegraphics[width=\hsize]{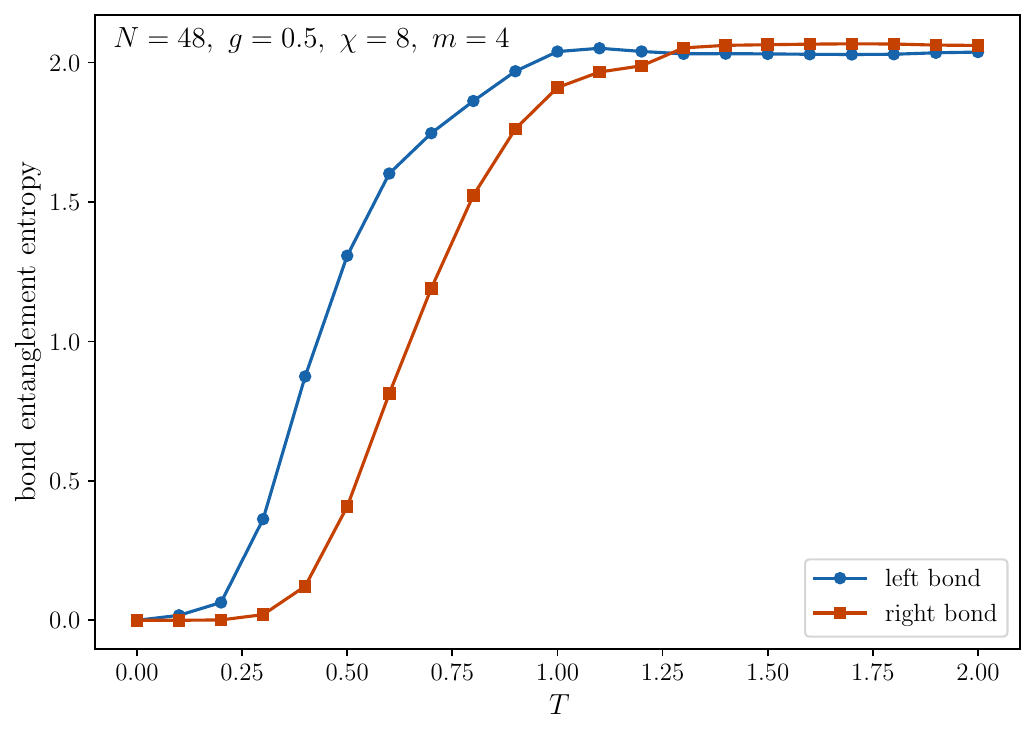}
    \caption{
        Temperature dependence of the entanglement entropy across the bipartition 
        at the left edge (blue circles) and the right edge (red squares), 
        for $N=48$, $g=0.5$, $\chi=8$, and $m=4$.
    }
\label{fig:bond_ent}
\end{figure}

Figure~\ref{fig:bond_ent} shows the entanglement entropy across bipartitions 
at the left and right edges of the system. 
At $T=0$, both quantities are zero, 
which indicates that the variational state is simply the ground state 
and does not rely on the auxiliary degrees of freedom. 
As the temperature increases, 
the bond entanglement also increases, 
demonstrating that the auxiliary sector is gradually activated to represent the finite-temperature MST state. 
The asymmetry between the left and right entanglement entropies reflects the intrinsic asymmetry of the Markov entropy.

%*%*%*%*%*%*%*%*%*%*%*%*%*%*%*%*%*%*%*%*%*%*%*%*%*%*%*%*%*%
%*%*%*%*%*%*%*%*%*%*%*%*%*%*%*%*%*%*%*%*%*%*%*%*%*%*%*%*%*%

%%%%%%%%%%%%%%%%%%%%%%%%%%%%%%%%%%%%%%%%%%%%
%%%%%%%%%%%%%% bibliography %%%%%%%%%%%%%%%%
%%%%%%%%%%%%%%%%%%%%%%%%%%%%%%%%%%%%%%%%%%%%
\nocite{apsrev42Control}
\bibliographystyle{apsrev4-2}
\bibliography{references}